\documentclass[11pt]{JHEP3}
\usepackage{amsfonts,amssymb,amsmath,amstext,amsbsy,amsthm,amscd,graphicx,epsfig,slashed,indentfirst,empheq}

\preprint{YITP-SB-14-47}

\title{Dynamics of Two-Dimensional $\CN=(2,2)$ Theories with Semichiral Superfields I}

\author{Jun Nian, Xinyu Zhang\\C. N. Yang Institute for Theoretical Physics, \\Stony Brook University, \\ Stony Brook, NY 11794-3840, USA\\ E-mail: \email{jnian@insti.physics.sunysb.edu}, \email{zhangxinyuphysics@gmail.com}}

\abstract{We analyze the dynamics of a general two-dimensional $\CN=(2,2)$ gauged linear sigma model with semichiral superfields. By computing the elliptic genera, we study the vacuum structure of the model. The result coincides with the model without using semichiral superfields. We also show that the low energy effective twisted superpotential contributed by semichiral superfields vanishes, whether we turn on twisted masses or not.}

\def\be{\begin{equation}}
\def\ee{\end{equation}}
\def\ben{\begin{eqnarray}}
\def\een{\end{eqnarray}}

\def\CN{\mathcal{N}}

\def\CW{\mathcal{W}}
\def\CY{\mathcal{Y}}

\def\hat{\widehat}
\def\tilde{\widetilde}
\def\bar{\overline}
\def\<{\langle}
\def\>{\rangle}

\def\U{\mathrm{U}}

\def\XXint#1#2#3{{\setbox0=\hbox{$#1{#2#3}{\int}$}
     \vcenter{\hbox{$#2#3$}}\kern-.5\wd0}}

\newcommand{\bb}{\mathbb}

\begin{document}

\section{Introduction}

There has been a lot of work on two-dimensional $\CN=(2,2)$ supersymmetric theories. For a review with numerous references, see~\cite{mirror}. Compared to the well-known two-dimensional $\CN=(2,2)$ chiral and twisted chiral superfields, semichiral superfields are less well studied in the literature.

Semichiral superfields were first introduced in Ref.~\cite{Buscher}. In Ref.~\cite{Off-shellSUSY} it was proved that to have a complete description of off-shell two-dimensional $\CN=(2,2)$ supersymmetry, one needs chiral, twisted chiral and semichiral superfields. However, except for some works, e.g. Ref.~\cite{Grisaru}, the majority of previous works on semichiral superfields focused on mathematical interpretations of sigma models at the classical level, while leaving many problems of quantum dynamics untouched. We shall fill this gap in a series of papers. As a first step, our goal in the present paper is not to study the most general two-dimensional $\CN=(2,2)$ supersymmetric theories with semichiral superfields. We will only consider a special type of models, namely the gauged linear sigma model (GLSM)~\cite{Witten-phases}. More general cases will be discussed in subsequent papers~\cite{followup}. We also limit ourselves to theories on a flat worldsheet. Theories on a sphere were studied in a separate paper~\cite{SemiLoc}.

The first elementary question of a supersymmetric model is whether supersymmetry is spontaneously broken or not. To answer this question, we need to compute the Witten index~\cite{Witten:1982df}, which gives the number of zero energy bosonic vacuum states minus the number of zero energy fermionic vacuum states. It is important because if supersymmetry is spontaneously broken then there are no zero energy ground states and the Witten index vanishes. It is also useful because it is a quasi-topological quantity, which depends only on F-terms and not on D-terms in the Lagrangian, and is exactly computable. In two dimensions, we can compute a more refined invariant, the elliptic genus, which can give more information about the vacuum structure of the theory. From the purely mathematical point of view, the elliptic genus of a sigma model captures important topological information of the target space.

Next, we want to go beyond the vacuum states. We assume that the vector fields used to gauge the semichiral superfields are ordinary vector fields, which can be equivalently organized using twisted chiral superfields.\footnote{Writing the vector multiplet as a twisted chiral superfield, the imaginary part of the highest component is the field strength of the vector field.} On the Coulomb branch, the gauge group $G$ is broken down to its Cartan subgroup $\U (1)^r$. In addition, we can turn on generic twisted masses for all the matter fields so that the matter fields become massive. At energies lower than all the mass scales in the theory, we can integrate out both W-bosons and matter fields, and the low energy effective theory is described by a model with only twisted chiral superfields. It is still beyond our scope to compute exactly the full low energy effective action. However, thanks to the special properties of supersymmetry, we can compute the effective twisted superpotential $\tilde{\CW}^{\rm eff}$ exactly. This quantity plays an essential role in the discussion of the sigma model/Landau-Ginzberg models correspondence~\cite{Witten-phases}, and determines the (twisted) chiral ring structure of the theory.

Recently, $\tilde{\CW}^{\rm eff}$ also appeared in the Bethe/gauge correspondence~\cite{Nekrasov:2009uh,Nekrasov:2009ui,Nekrasov:2014xaa}. In this remarkable correspondence, $\tilde{\CW}^{\rm eff}$ computed from a two-dimensional $\CN=(2,2)$ supersymmetric gauge theory is conjectured to be identified with the Yang-Yang function $\CY$ of a quantum integrable system. In the previous discussions, the matter multiplets are always built using chiral superfields. It is natural to ask whether semichiral superfields can give new contributions.

The organization of this paper is as follows. In section~\ref{sec:Semi} we review the two-dimensional $\CN=(2,2)$ supersymmetric GLSMs with semichiral superfields. In section~\ref{sec:EG} we compute the elliptic genus. After a general discussion, we work out two important examples, namely the elliptic genus of the Eguchi-Hanson space and the Taub-NUT space. The sigma models built from semichiral superfields give exactly the same results as those without using the semichiral superfields. In section~\ref{sec:LEETS} we compute the low energy effective twisted superpotential. We find that the contribution from the semichiral superfield vanishes, even if we turn on generic twisted masses. Finally, in section~\ref{sec:con} we give a conclusion and discuss some possible directions for future work. Since computations of semichiral superfields are usually unavoidably lengthy, we put some details in the appendices.

\section{Two-dimensional $\CN=(2,2)$ supersymmetry}\label{sec:Semi}

In this section, we review two-dimensional $\CN=(2,2)$ supersymmetric theories in order to be more self-contained. The natural language to describe the theories we will study in a compact form is the two-dimensional $\CN=(2,2)$ superspace (see Appendix~\ref{App:superspace}). Some detailed formulae written in components are collected in Appendix~\ref{App:GLSMsemi}.

\subsection{Supersymmetric multiplets}

Using the two-dimensional $\CN=(2,2)$ superspace, we can describe all possible choices of superfields which can appear in a general supersymmetric model. We will be brief in the discussion of better understood multiplets and focus on semichiral multiplets.

The basic matter multiplet is described by a chiral superfield $\Phi$, which contains a scalar $\phi$, a fermion $\psi_{\pm \Phi}$ and an auxiliary field $F_{\Phi}$.~\footnote{Here we add a subscript $\Phi$ to distinguish them from components of semichiral superfields.} It is defined by the condition
\be
\bar{\mathbb{D}}_\pm \Phi = 0 \, .
\ee
Similarly we can define its conjugate to be an anti-chiral superfield $\bar{\Phi}$. Indeed, chiral superfields can be obtained by dimensional reduction from $\CN=1$ chiral superfields in four dimensions.

The basic vector multiplet contains a real gauge field $A_\mu$, a complex scalar $\hat{\sigma}$, two Weyl fermion $\lambda_\pm$ and an auxiliary real scalar $D$. It can also be obtained via dimensional reduction from four-dimensional vector multiplet.

Although the $\CN=(2,2)$ supersymmetry algebra in two dimensions can be obtained by dimensional reduction from $\CN=1$ supersymmetry algebra in four dimensions, not all superfields in two dimensions can be obtained simply via dimensional reduction from four dimensions. An important superfield which is unique in two dimensions is the twisted chiral superfield $\Sigma$, defined by the conditions
\be
\bar{\mathbb{D}}_+ \Sigma = {\mathbb{D}}_- \Sigma =  0 \, .
\ee
Similarly we can define a twisted anti-chiral superfield $\bar{\Sigma}$. The components of a vector multiplet can be reorganized into a twisted chiral superfield $\Sigma$ in the following way:
\ben
  \Sigma | &=& \hat{\sigma}\, ,\quad \mathbb{D}_+ \Sigma | = i \bar{\lambda}_+\, ,\quad \bar{\mathbb{D}}_- \Sigma | = i \lambda_-\, ,\quad \bar{\mathbb{D}}_- \mathbb{D}_+ \Sigma | = D - i F_{01}\, , \nonumber\\~
  \bar{\Sigma} | &=& \bar{\hat{\sigma}}\, ,\quad \bar{\mathbb{D}}_+ \bar{\Sigma} | = - i \lambda_+\, ,\quad \mathbb{D}_- \bar{\Sigma} | = - i \bar{\lambda}_-\, ,\quad \mathbb{D}_- \bar{\mathbb{D}}_+ \bar{\Sigma} | = - (D + i F_{01})\, .
\een

When we construct models with gauge fields, it is more convenient to use the covariant approach, in which the gauge connections are incorporated in the supercovariant derivatives. Accordingly, the anticommutation relations of the supercovariant derivatives are modified as follows
\be
    \{\mathcal{D}_\pm,\, \bar{\mathcal{D}}_\pm \} = -2 i D_\pm\, ,\quad \{ \bar{\mathcal{D}}_+,\, \mathcal{D}_- \} \equiv \Sigma\, ,\quad \{\mathcal{D}_+,\, \bar{\mathcal{D}}_- \} \equiv \bar{\Sigma}\, ,
\ee
where $D_\pm \equiv \partial_\pm + A_\pm$ is the gauge covariant derivative, and $\Sigma$ is the field strength superfield, which is twisted chiral.

However, this is not the end of the story. The final building block is semichiral superfields. The left-semichiral and the right-semichiral multiplets are define by
  \be
    \bar{\mathbb{D}}_+ \bb{X}_L = 0\, ,\quad \bar{\mathbb{D}}_- \bb{X}_R = 0\, ,
  \ee
and similarly, we have for their conjugates
  \be
    \mathbb{D}_+ \overline{\bb{X}}_L = 0\, ,\quad \mathbb{D}_- \overline{\bb{X}}_R = 0\, .
  \ee
In order to have a better understanding of these semichiral superfields, we will expand the superfields and write down their components. It is convenient to treat the left-semichiral and the right-semichiral multiplets simultaneously by imposing a weaker constraint:
\be
  \bar{\mathbb{D}}_+ \bar{\mathbb{D}}_- \mathbb{X} = 0\, ,\quad \mathbb{D}_+ \mathbb{D}_- \bar{\mathbb{X}} = 0\, .
\ee
Then we define the components
\begin{align}
    X & = \mathbb{X} |\, ,\quad \psi_\pm \equiv \mathbb{D}_\pm \mathbb{X} |\, ,\quad \bar{\chi}_\pm = \bar{\mathbb{D}}_\pm \mathbb{X} |\, , \quad F \equiv \mathbb{D}_+ \mathbb{D}_- \mathbb{X} |\, ,  \nonumber\\~
    \quad M_{-+} & = \mathbb{D}_+ \bar{\mathbb{D}}_- \mathbb{X} |\, ,\quad M_{+-} = \mathbb{D}_- \bar{\mathbb{D}}_+ \mathbb{X} |\, ,\quad M_{\pm \pm} = \mathbb{D}_\pm \bar{\mathbb{D}}_\pm \mathbb{X} |\, ,\quad \bar{\eta}_\pm = \mathbb{D}_+ \mathbb{D}_- \bar{\mathbb{D}}_\pm \mathbb{X} |\, ,
\end{align}
and
\begin{align}
    \bar{X} & = \bar{\mathbb{X}} |\, ,\quad \bar{\psi}_\pm = \bar{\mathbb{D}}_\pm \bar{\mathbb{X}} |\, ,\quad \chi_\pm = \mathbb{D}_\pm \bar{\mathbb{X}} |\, , \quad \bar{F} = \bar{\mathbb{D}}_+ \bar{\mathbb{D}}_- \bar{\mathbb{X}} |\, , \nonumber\\~
 \bar{M}_{-+} & = \bar{\mathbb{D}}_+ \mathbb{D}_- \bar{\mathbb{X}} |\, ,\quad \bar{M}_{+-} = \bar{\mathbb{D}}_- \mathbb{D}_+ \bar{\mathbb{X}} |\, ,\quad \bar{M}_{\pm \pm} = \bar{\mathbb{D}}_\pm \mathbb{D}_\pm \bar{\mathbb{X}} |\, ,\quad \eta = \bar{\mathbb{D}}_+ \bar{\mathbb{D}}_- \mathbb{D}_\pm \bar{\mathbb{X}} |\, .
\end{align}
We then impose the constraints for each multiplet, and some component fields should vanish,
\begin{align}
    \textrm{For} \,\, \mathbb{X}_L: & \quad \chi_+ = M_{+-} = M_{++} = \eta_+ = 0\, ,\nonumber\\
    \textrm{For} \,\, \mathbb{X}_R: & \quad \chi_- = M_{-+} = M_{--} = \eta_- = 0\, ,\nonumber\\
    \textrm{For} \,\, \bar{\mathbb{X}}_L: & \quad \bar{\chi}_+ = \bar{M}_{+-} = \bar{M}_{++} = \bar{\eta}_+ = 0\, ,\nonumber\\
    \textrm{For} \,\, \bar{\mathbb{X}}_R: & \quad \bar{\chi}_- = \bar{M}_{-+} = \bar{M}_{--} = \bar{\eta}_- = 0\, .
\end{align}

\subsection{Gauged linear sigma models}

It was shown in Ref.~\cite{Off-shellSUSY} that the most general two-dimensional $\mathcal{N} = (2, 2)$ GLSM can be constructed using chiral, twisted chiral and semichiral superfields. The GLSM with chiral and twisted chiral superfields has been exploited at length in the literature. Hence, we will focus on the indispensable but poorly understood building block, the action with semichiral superfields. To obtain a gauged linear linear model with physical kinetic terms, one needs both left-semichiral and right-semichiral superfields simultaneously. Models with only left-semichiral or only right-semichiral superfields turn out to be topological.

In this paper, we gauge the semichiral superfields using the constrained semichiral vector multiplets \cite{MartinMarcos}, which will be reviewed in the following. Let us first discuss using the semichiral vector multiplet to gauge the semichiral superfields. Here we only consider the abelian case, and the nonabelian case is discussed in Ref.~\cite{nonAbelian}. An abelian semichiral vector multiplet can be described by three real vector superfields $(V_L,\, V_R,\, V')$~\cite{SVM-1}. If we define
  \be
    \mathbb{V} \equiv \frac{1}{2} (- V' + i (V_L - V_R))\, ,\quad \widetilde{\mathbb{V}} \equiv \frac{1}{2} (- V' + i (V_L + V_R))\, ,
  \ee
the action for a pair of semichiral superfields is
  \be
    S = \int d^2 x\, d^4 \theta\, K\, ,
  \ee
  where
  \be
    K = \overline{\bb{X}}_L e^{V_L} \bb{X}_L + \overline{\bb{X}}_R e^{V_R} \bb{X}_R + \alpha \overline{\bb{X}}_L e^{i \bar{\tilde{\mathbb{V}}}} \bb{X}_R + \alpha \overline{\bb{X}}_R e^{-i \tilde{\mathbb{V}}} \bb{X}_L\, ,\label{GLSMsemiSVM}
  \ee
  with $|\alpha| > 1$. This action is invariant under the gauge transformations:
  \be\label{eq:SVMgauge-1}
    \delta \mathbb{X}_L = e^{i \Lambda_L} \mathbb{X}_L\, ,\quad \delta \mathbb{X}_R = e^{i \Lambda_R} \mathbb{X}_R\, ,
  \ee
  \be\label{eq:SVMgauge-2}
    \delta V_L = i (\bar{\Lambda}_L - \Lambda_L)\, ,\quad \delta V_R = i (\bar{\Lambda}_R - \Lambda_R)\, ,\quad \delta V' = \Lambda_R + \bar{\Lambda}_R - \Lambda_L - \bar{\Lambda}_L\, ,
  \ee
or equivalently,
  \be\label{eq:SVMgauge-3}
    \delta \mathbb{V} = \Lambda_L - \Lambda_R\, ,\quad \delta \widetilde{\mathbb{V}} = \Lambda_L - \bar{\Lambda}_R\, .
  \ee

To review the constrained semichiral vector multiplet, we first see that one can define two independent gauge invariant field strengths for the semichiral vector multiplet
  \be
    \mathbb{F} \equiv \bar{\mathbb{D}}_+ \bar{\mathbb{D}}_- \mathbb{V}\, ,\quad \widetilde{\mathbb{F}} \equiv \bar{\mathbb{D}}_+ \mathbb{D}_- \widetilde{\mathbb{V}}\, .
  \ee
Here $\mathbb{F}$ is a chiral superfield, and $\widetilde{\mathbb{F}}$ is a twisted chiral superfield. The constrained semichiral vector multiplet can be viewed as a semichiral vector multiplet \cite{MartinMarcos} with an additional term:
\be
  \int d^2 \theta \, \hat{\Phi} \mathbb{F} + c.c.\, ,
\ee
where $\hat{\Phi}$ is a chiral Lagrange multiplier, and it imposes the constraint
\be
  \mathbb{F} = 0\, .
\ee
Since this additional term is a F-term, which is SUSY exact, it does not affect the result of localization, as long as it does not introduce some additional constraints for instance on the R-charges. Therefore, in many cases we can use the constrained semichiral vector multiplet to replace the vector multiplet without changing the result of localization.

We can perform a partial gauge fixing $V'=V_L-V_R=0$; this leaves just a chiral gauge invariance as a residual gauge invariance. The theory given in Eq.~\eqref{GLSMsemiSVM} then becomes
  \be
    K = \overline{\bb{X}}_L e^V \bb{X}_L + \overline{\bb{X}}_R e^V \bb{X}_R + \alpha \overline{\bb{X}}_L e^V \bb{X}_R + \alpha \overline{\bb{X}}_R e^V \bb{X}_L\, ,\label{GLSMsemi}
  \ee
The vector superfield $V$ can be viewed as a constrained semichiral vector multiplet after partially gauge fixing the full semichiral gauge freedom, and $\mathbb{X}_L$ and $\mathbb{X}_R$ have the same gauge charge. We can expand the action into component fields. The result is quite lengthy, and is written down in Appendix~\ref{App:GLSMsemi} for interested readers. To summarize, it is more natural to use the semichiral vector multiplet to gauge the semichiral multiplets, but after a partial gauge fixing the constrained semichiral vector multiplet is equivalent to an ordinary vector multiplet. In this sense, one can also use the ordinary vector multiplet to gauge the semichiral multiplets. Hence, the ordinary vector multiplet is a special case of the constrained semichiral vector multiplet.

Now we generalize the above discussion to a theory with gauge group $\U(1)^N$ and $N_F$ flavors. One can turn on twisted mass deformations for the model with flavor symmetry. To add the most general twisted masses, one first gauges the flavor symmetry using the semichiral vector superfield and then sets the scalar component of the semichiral vector superfield to be a nonzero constant value, and finally requires the other components of the semichiral vector superfield vanish. Supersymmetry is not broken by twisted masses. If we write explicit color and flavor indices, we have
  \begin{align}
    K & = \bar{\mathbb{X}}^L_{a, i} \left(e^V \right)^{ab} \left(e^{V_L}\right)^{ij} \mathbb{X}^L_{b, j} + \bar{\mathbb{X}}^R_{a, i} \left(e^V \right)^{ab} \left(e^{V_R}\right)^{ij} \mathbb{X}^R_{b, j} \nonumber\\
   {} & \quad + \alpha \left[ \bar{\mathbb{X}}^L_{a, i} \left(e^V \right)^{ab} \left(e^{i \bar{\widetilde{\mathbb{V}}}} \right)^{ij} \mathbb{X}^R_{b, j} + \bar{\mathbb{X}}^R_{a, i} \left(e^V \right)^{ab} \left(e^{-i \widetilde{\mathbb{V}}} \right)^{ij} \mathbb{X}^L_{b, j} \right]\, ,
  \end{align}
  where $a, b = 1, \cdots, N$, $i, j = 1, \cdots, N_F$.

\section{Elliptic genus}\label{sec:EG}

The elliptic genus can be computed both using the Hamiltonian formalism~\cite{Witten} and the path integral formalism~\cite{Gadde-Gukov, Benini-genus-1, Benini-genus-2}. In this section, we will compute the elliptic genus of the GLSM with semichiral superfields using both methods, and our discussion will be restricted to the Abelian GLSM.

\subsection{Hamiltonian formalism}

The elliptic genus is defined in the Hamiltonian formalism as a refined Witten index,
  \be\label{eq:DefEG}
    Z = \textrm{Tr}_{\rm RR} (-1)^F q^{H_L} \bar{q}^{H_R} y^J \prod_a x_a^{K_a}\, ,
  \ee
where the trace is taken in the RR sector, in which fermions have periodic boundary conditions, and $F$ is the fermion number. In Euclidean signature, $H_L =\frac{1}{2} (H + i P)$ and $H_R =\frac{1}{2} (H - i P)$ are the left- and the right-moving Hamiltonians. $J$ and $K_a$ are the R-symmetry and the $a$-th flavor symmetry generators, respectively. It is standard to also define
  \be
    q \equiv e^{2\pi i \tau}\, ,\quad x_a \equiv e^{2\pi i u_a}\, ,\quad y \equiv e^{2\pi i z}\, .
  \ee
If $u_a=z=0$ the elliptic genus reduces to the Witten index, and computes the Euler characteristic of the target space if there is a well-defined geometric description.

The contributions from different multiplets can be computed independently, and we will only consider the unexplored contribution from the semichiral multiplet. As we have seen in Appendix~\ref{App:GLSMsemi}, the physical component fields of the semichiral superfield $\mathbb{X}$ are two complex scalars $X_L$ and $X_R$, and spinors $\psi_\pm'$, $\chi_-^L$ and $\chi_+^R$. All fields have the same flavor symmetry charge $Q$. The $R$-charges of $(X_L, X_R, \psi_+', \psi_-' \chi_-^L,\chi_+^R)$ are $(\frac{R}{2},\frac{R}{2}, \frac{R}{2}-1, \frac{R}{2}, \frac{R}{2}, \frac{R}{2}+1)$.

Let us consider the fermionic zero modes first. We denote the zero modes of $\psi_+'$ and $\bar{\psi}_+'$ as $\psi_{+, 0}'$ and $\bar{\psi}_{+, 0}'$, respectively. They satisfy
  \be
    \{\psi_{+, 0}'\, ,\, \bar{\psi}_{+, 0}' \} = 1\, ,
  \ee
  which can be represented in the space spanned by $|\downarrow \rangle$ and $|\uparrow \rangle$ with
  \be
    \psi_{+, 0}'\, |\downarrow \rangle = |\uparrow \rangle\, ,\quad \bar{\psi}_{+, 0}'\, |\uparrow \rangle = |\downarrow \rangle\, .
  \ee
  One of $|\downarrow \rangle$ and $|\uparrow \rangle$ can be chosen to be bosonic, while the other is fermionic.  Under the $U(1)_R$ the zero modes transform as
  \be
    \psi_{+, 0}' \rightarrow e^{-i \pi z (\frac{R}{2} - 1) } \psi_{-, 0}'\, ,\quad \bar{\psi}_{+, 0}' \rightarrow e^{i \pi z (\frac{R}{2} - 1) } \bar{\psi}_{-, 0}'\, ,
  \ee
  while under $U(1)_f$ they transform as
  \be
    \psi_{+, 0}' \rightarrow e^{-i \pi u Q } \psi_{-, 0}'\, ,\quad \bar{\psi}_{+, 0}' \rightarrow e^{i \pi u Q } \bar{\psi}_{-, 0}'\, ,
  \ee
  These two states contribute a factor
  \be
    e^{-i \pi z (\frac{R}{2} - 1) }\, e^{-i \pi u Q } - e^{i \pi z (\frac{R}{2} - 1) }\, e^{i \pi u Q }
  \ee
  to the elliptic genus. Similarly, the contributions of the other zero modes are
  \ben
    &&(\psi_{-, 0}', \bar{\psi}_{-, 0}'): \quad e^{i \pi z \frac{R}{2}}\, e^{i \pi u Q } - e^{-i \pi z \frac{R}{2}}\, e^{-i \pi u Q } \, , \nonumber\\~
    &&(\chi_{-, 0}^L, \bar{\chi}_{-, 0}^L): \quad  e^{i \pi z \frac{R}{2}}\, e^{i \pi u Q } - e^{-i \pi z \frac{R}{2}}\, e^{-i \pi u Q }\, ,\nonumber\\~
    &&(\chi_{+,0}^R, \bar{\chi}_{+,0}^R): \quad e^{-i \pi z (\frac{R}{2} + 1) }\, e^{-i \pi u Q } - e^{i \pi z (\frac{R}{2} + 1) }\, e^{i \pi u Q }\, .
  \een
  The contributions from the bosonic zero modes are relatively simple. They are
  \be
    \frac{1}{\left[ \left(1 - e^{i \pi z \frac{R}{2}}\, e^{i \pi u Q }\right)\cdot \left(1 - e^{-i \pi z \frac{R}{2}}\, e^{-i \pi u Q } \right)\right]^2} \, .
  \ee
  Bringing all the factors together, we obtain the zero mode part of the elliptic genus:
  \be\label{eq:zeromodes}
    \frac{\left[1 - e^{i \pi (R - 2) z}\, e^{2 i \pi u Q } \right]\cdot \left[1 - e^{i \pi (R + 2) z}\, e^{2 i \pi u Q } \right]}{\left(1 - e^{i \pi R z}\, e^{2 i \pi u Q } \right)^2} = \frac{\left(1 - y^{\frac{R}{2} - 1} x^Q \right)\cdot \left(1 - y^{\frac{R}{2} + 1} x^Q\right)}{(1 - y^{\frac{R}{2}} x^Q)^2} \, .
  \ee

  We then consider the nonzero modes. The contribution from the fermionic sector $(\psi_\pm', \chi_-^L, \chi_+^R)$ is
  \begin{align}
    \prod_{n=1}^\infty & \left(1 - q^n e^{2 i \pi z (\frac{R}{2} - 1)}\, e^{2 i \pi u Q} \right) \cdot \left(1 - q^n e^{-2 i \pi z (\frac{R}{2} - 1)}\, e^{-2 i \pi u Q} \right) \nonumber\\
    {} & \cdot \left(1 - \bar{q}^n e^{2 i \pi z \frac{R}{2}}\, e^{2 i \pi u Q} \right) \cdot \left(1 - \bar{q}^n e^{-2 i \pi z \frac{R}{2}}\, e^{-2 i \pi u Q} \right) \nonumber\\
    {} & \cdot \left(1 - q^n e^{2 i \pi z (\frac{R}{2} + 1)}\, e^{2 i \pi u Q} \right) \cdot \left(1 - q^n e^{-2 i \pi z (\frac{R}{2} + 1)}\, e^{-2 i \pi u Q} \right) \nonumber\\
    {} & \cdot \left(1 - \bar{q}^n e^{2 i \pi z \frac{R}{2}}\, e^{2 i \pi u Q} \right) \cdot \left(1 - \bar{q}^n e^{-2 i \pi z \frac{R}{2}}\, e^{-2 i \pi u Q} \right)\, ,
  \end{align}
  while the contribution from the bosonic sector ($X_L, X_R$) is
  \begin{align}
    \prod_{n=1}^\infty & \frac{1}{\left(1 - q^n e^{2 i \pi z \frac{R}{2}}\, e^{2 i \pi u Q} \right) \cdot \left(1 - q^n e^{-2 i \pi z \frac{R}{2}}\, e^{-2 i \pi u Q} \right)} \nonumber\\
    {} & \cdot \frac{1}{\left(1 - \bar{q}^n e^{2 i \pi z \frac{R}{2}}\, e^{2 i \pi u Q} \right) \cdot \left(1 - \bar{q}^n e^{-2 i \pi z \frac{R}{2}}\, e^{-2 i \pi u Q} \right)} \nonumber\\
    {} & \cdot \frac{1}{\left(1 - q^n e^{2 i \pi z \frac{R}{2}}\, e^{2 i \pi u Q} \right) \cdot \left(1 - q^n e^{-2 i \pi z \frac{R}{2}}\, e^{-2 i \pi u Q} \right)} \nonumber\\
    {} & \cdot \frac{1}{\left(1 - \bar{q}^n e^{2 i \pi z \frac{R}{2}}\, e^{2 i \pi u Q} \right) \cdot \left(1 - \bar{q}^n e^{-2 i \pi z \frac{R}{2}}\, e^{-2 i \pi u Q} \right)}\, .
  \end{align}
  Hence, the nonzero modes contribute to the elliptic genus a factor
  \begin{align}
    \quad \prod_{n=1}^\infty & \frac{\left(1 - q^n e^{2 i \pi z (\frac{R}{2} - 1)}\, e^{2 i \pi u Q} \right) \cdot \left(1 - q^n e^{-2 i \pi z (\frac{R}{2} - 1)}\, e^{-2 i \pi u Q} \right)}{\left(1 - q^n e^{2 i \pi z \frac{R}{2}}\, e^{2 i \pi u Q} \right) \cdot \left(1 - q^n e^{-2 i \pi z \frac{R}{2}}\, e^{-2 i \pi u Q} \right)} \nonumber\\
    {} & \cdot \frac{\left(1 - q^n e^{2 i \pi z (\frac{R}{2} + 1)}\, e^{2 i \pi u Q} \right) \cdot \left(1 - q^n e^{-2 i \pi z (\frac{R}{2} + 1)}\, e^{-2 i \pi u Q} \right)}{\left(1 - q^n e^{2 i \pi z \frac{R}{2}}\, e^{2 i \pi u Q} \right) \cdot \left(1 - q^n e^{-2 i \pi z \frac{R}{2}}\, e^{-2 i \pi u Q} \right)} \nonumber\\
    = \prod_{n=1}^\infty & \frac{\left(1 - q^n y^{\frac{R}{2} - 1} x^Q \right) \cdot \left(1 - q^n (y^{\frac{R}{2} - 1} x^Q)^{-1} \right)}{\left(1 - q^n y^{\frac{R}{2}} x^Q \right) \cdot \left(1 - q^n (y^{\frac{R}{2}} x^Q)^{-1} \right)} \cdot \frac{\left(1 - q^n y^{\frac{R}{2} + 1} x^Q \right) \cdot \left(1 - q^n (y^{\frac{R}{2} + 1} x^Q)^{-1} \right)}{\left(1 - q^n y^{\frac{R}{2}} x^Q \right) \cdot \left(1 - q^n (y^{\frac{R}{2}} x^Q)^{-1} \right)}\, .\label{eq:nonzeromodes}
  \end{align}

  Taking both the zero modes \eqref{eq:zeromodes} and the nonzero modes \eqref{eq:nonzeromodes} into account, we obtain
  \begin{align}
    {} & \frac{\left(1 - y^{\frac{R}{2} - 1} x^Q \right)\cdot \left(1 - y^{\frac{R}{2} + 1} x^Q\right)}{(1 - y^{\frac{R}{2}} x^Q)^2} \cdot \prod_{n=1}^\infty \frac{\left(1 - q^n y^{\frac{R}{2} - 1} x^Q \right) \cdot \left(1 - q^n (y^{\frac{R}{2} - 1} x^Q)^{-1} \right)}{\left(1 - q^n y^{\frac{R}{2}} x^Q \right) \cdot \left(1 - q^n (y^{\frac{R}{2}} x^Q)^{-1} \right)} \nonumber\\
    {} & \cdot \prod_{n=1}^\infty \frac{\left(1 - q^n y^{\frac{R}{2} + 1} x^Q \right) \cdot \left(1 - q^n (y^{\frac{R}{2} + 1} x^Q)^{-1} \right)}{\left(1 - q^n y^{\frac{R}{2}} x^Q \right) \cdot \left(1 - q^n (y^{\frac{R}{2}} x^Q)^{-1} \right)}\, .\label{eq:EGresultHam}
  \end{align}
Using the formula
  \be
    \vartheta_1 (\tau, z) = -i y^{1/2} q^{1/8} \prod_{n=1}^\infty (1 - q^n) \prod_{n=0}^\infty (1 - y q^{n+1}) (1 - y^{-1} q^n)\, ,
  \ee
where
\be
  q \equiv e^{2 \pi i \tau}\, ,\quad y \equiv e^{2 \pi i z}\, ,
\ee
we can rewrite (\ref{eq:EGresultHam}) as
  \be
    Z^{1-loop} (\tau,\, u,\, z) = \frac{\vartheta_1 (\tau,\, z \left(\frac{R}{2} + 1\right) + u Q)}{\vartheta_1 (\tau,\, z \frac{R}{2} + u Q)} \cdot \frac{\vartheta_1 (\tau,\, z \left(\frac{R}{2} - 1\right) + u Q)}{\vartheta_1 (q,\, z \frac{R}{2} + u Q)}\, .
  \ee
Comparing to the contribution of a chiral superfield \cite{Gadde-Gukov, Benini-genus-1}, we see that the 1-loop determinant of the elliptic genus for one pair of semichiral superfields is equal to the product of the 1-loop determinants for two chiral superfields with the opposite R-charge and the opposite flavor charge, which is consistent with the result of the semichiral gauged linear sigma model localized on the two-sphere \cite{SemiLoc}.

\subsection{Path integral formalism}

The elliptic genus can be equivalently described in the path integral formalism as a twisted partition function on the torus, we may apply the technique of localization to compute it.

Recall that the Witten index is expressed in the path integral formalism as the partition function of the theory on a torus, with periodic boundary conditions for both bosons and fermions. To deform the Witten index into the elliptic genus, we should specify twisted boundary conditions for all fields. Equivalently, we can keep the periodic boundary conditions and introduce background gauge fields $A^R$ and $A^{f, a}$ for the R-symmetry and the $a$-th flavor-symmetry, respectively. They are related to the parameters in the definition of elliptic genus via
\be\label{eq:fugacity}
    z \equiv \oint A_1^R\, d x_1- \tau \oint A_2^R\, d x_2\, , \quad u_a \equiv \oint A_1^{f, a} d x_1 - \tau \oint A_2^{f, a} d x_2\,  .
\ee

Following the general principle of localization, if we regard the background gauge fields as parameters in the theory, we only need the free part of the Lagrangian in order to compute the elliptic genus. The free part of the Lagrangian in the Euclidean signature is
  \begin{align}
    \mathcal{L}^{\rm free} & = D_\mu \bar{X}^I D^\mu X_I + i \bar{X}^I D X_I + \bar{F}^I F_I - \bar{M}^{++, I} M_{++, I} - \bar{M}^{--, I} M_{--, I} - \bar{M}^{+-, I} M_{-+, I} - \bar{M}^{-+, I} M_{+-, I} \nonumber\\
    {} & \quad - \bar{M}^{++, I} (-2 i D_+ X_I) - \bar{M}^{--, I} 2 i D_- X_I + \bar{X}^I (-2 i D_+ M^{++}_I) + \bar{X}^I (2 i D_- M^{--}_I) \nonumber\\
    {} & \quad - i \bar{\psi}^I \gamma^\mu D_\mu \psi_I - \bar{\eta}^I \psi_I - \bar{\psi}^I \eta_I + i \bar{\chi}^I \gamma^\mu D_\mu \chi_I\, ,
  \end{align}
  where the covariant derivative is defined as
  \be
    D_\mu \equiv \partial_\mu - \hat{Q} u_\mu - \hat{R} z_\mu\, ,
  \ee
  and the operators $\hat{Q}$ and $\hat{R}$ acting on different fields give their corresponding $U(1)_f$ and $U(1)_R$ charges as follows:
  \be
    \begin{array}{c||c|c|c|c|c|c|c|c|c|c|c|c}
      {} & X & \psi_+ & \psi_- & F & \chi_+ & \chi_- & M_{++} & M_{--} & M_{+-} & M_{-+} & \eta_+ & \eta_-\\
      \hline
      \hat{Q} & Q & Q & Q & Q & Q & Q & Q & Q & Q & Q & Q & Q\\
      \hline
      \hat{R} & \frac{R}{2} & \frac{R}{2}-1 & \frac{R}{2} & \frac{R}{2}-1 & \frac{R}{2}+1 & \frac{R}{2} & \frac{R}{2} & \frac{R}{2} & \frac{R}{2}+1 & \frac{R}{2}-1 & \frac{R}{2} & \frac{R}{2}-1
    \end{array}
  \ee

  The BPS equations are obtained by setting the SUSY transformations of fermions to zero. The solutions to the BPS equtions provide the background that can perserve certain amount of supersymmetry. In this case, the BPS equations have only trivial solutions, i.e., all the fields in the semichiral multiplets are vanishing.

  We adopt the metric on the torus
  \be
    ds^2 = g_{ij}\, dx^i\, dx^j\, ,
  \ee
  where
  \be
    g_{ij} = \frac{1}{\tau_2}
    \left(\begin{array}{cc}
      1 & \tau_1\\
      \tau_1 & |\tau|^2
    \end{array}\right)\, ,
  \ee
  and $\tau = \tau_1 + i \tau_2$ is the complex structure, and we expand all the fields in the modes
  \begin{displaymath}
    e^{2 \pi i (n x_1 - m x_2)}\, ,
  \end{displaymath}
  where $n,\, m \in \mathbb{Z}$. Then we can integrate out the auxiliary fields, and calculate the 1-loop determinant of the free part of the Lagrangian on the torus. The result is
  \be
    Z^{1-loop} = \prod_{m,\, n \in \mathbb{Z}} \frac{\left( m + n \tau - Q u - (\frac{R}{2} + 1) z \right) \cdot \left( m + n \tau - Q u - (\frac{R}{2} - 1) z \right)}{\left( m + n \tau - (Q u + \frac{R}{2} z) \right) \cdot \left( m + n \tau - (Q u + \frac{R}{2} z) \right)} \, .
  \ee
  After regularization, this expression can be written in terms of theta functions:
  \be
    Z^{1-loop} (\tau,\, u,\, z) = \frac{\vartheta_1 (\tau,\, z \left(\frac{R}{2} + 1\right) + u Q)}{\vartheta_1 (\tau,\, z \frac{R}{2} + u Q)} \cdot \frac{\vartheta_1 (\tau,\, z \left(\frac{R}{2} - 1\right) + u Q)}{\vartheta_1 (q,\, z \frac{R}{2} + u Q)}\, .
  \ee
Using the localization technique, Refs.~\cite{Benini-genus-1, Benini-genus-2} have shown that for a large class of 2-dimensional $\mathcal{N}=(0,2)$ GLSM's the elliptic genus is given by
\be
Z = \frac{1}{|W|} \sum_{u_* \in \mathfrak{M}_{\textrm{sing}}^*} \textrm{JK-Res}_{u_*} (Q(u_*), \eta)\, Z_{1-loop} (u)\, ,
\ee
where $u$ is the holonomy of the gauge field on the spacetime torus $T^2$:
\be
  u \equiv \oint A_t \, dt - \tau \oint A_s \, ds \quad (t, s: \textrm{temporal and spatial direction})
\ee
which is different from the fugacities $u_a$ for the flavor symmetries defined in Eq.~\eqref{eq:fugacity}. As shown in Refs.~\cite{Benini-genus-1, Benini-genus-2}, the final result is the Jeffrey-Kirwan residue (see also Appendix~\ref{App:JK}), which was discussed in Refs.~\cite{JK-1, JK-2} and more recently in Refs.~\cite{Benini-genus-1, Benini-genus-2, Lee}.

\subsection{Eguchi-Hanson space}
Eguchi-Hanson space is the simplest example of the ALE spaces, and can be constructed via hyperk\"ahler quotient in terms of semichiral superfields \cite{Marcos}:
\begin{align}
  {\cal L} & = -\frac{1}{2 e^2} \int d^4 \theta (\bar{\tilde{\mathbb{F}}} \tilde{\mathbb{F}} - \bar{\mathbb{F}} \mathbb{F}) + \left(i \int d^2 \theta \, \Phi\, \mathbb{F} + c.c. \right) + \left( i \int d^2 \tilde{\theta}\, t\, \tilde{\mathbb{F}} + c.c.\right) \nonumber\\~
  & \quad - \int d^4 \theta \left[\bar{\mathbb{X}}_i^L\, e^{Q_i V_L}\, \mathbb{X}_i^L + \bar{\mathbb{X}}_i^R\, e^{Q_i V_R}\, \mathbb{X}_i^R + \alpha (\bar{\mathbb{X}}_i^L\, e^{i Q_i \bar{\tilde{\mathbb{V}}}}\, \mathbb{X}_i^R + \bar{\mathbb{X}}_i^R\, e^{-i Q_i \tilde{\mathbb{V}}}\, \mathbb{X}_i^L )  \right]\, ,\label{eq:semi-EH}
\end{align}
where $i = 1, 2$, and for simplicity we set $t=0$.

The model \eqref{eq:semi-EH} has $\mathcal{N} = (4, 4)$ supersymmetry, and the R-symmetry is $SO(4) \times SU(2) \cong SU(2)_1 \times SU(2)_2 \times SU(2)_3$ \cite{Lee}. Hence, we can assign the R-charges $(Q_1, Q_2, Q_R)$, where $Q_R$ corresponds to the $U(1)_R$ charge that we discussed in the previous section. Similar to Ref.~\cite{Lee}, we choose the supercharges $\mathcal{Q}_-$ and $\mathcal{Q}_+$ to be in the representation $(2, 2, 1)$ and $(2, 1, 2)$ respectively under the R-symmetry group. Moreover, the flavor symmetry $Q_f$ now becomes $SU(2)_f$. In this case, the fields appearing in the model \eqref{eq:semi-EH}, which are relevant for the elliptic genus, have the following charge assignments:
  \be\label{charge-1}
    \begin{array}{c||c|c|c|c|c|c||c|c|c|c|c|c}
      {} & X_1^L & X_1^R & \psi_{1+}^{(2)} & \psi_{1-}^{(2)} & \chi_{1+}^R & \chi_{1-}^L & X_2^L & X_2^R & \psi_{2+}^{(2)} & \psi_{2-}^{(2)} & \chi_{2+}^R & \chi_{2-}^L\\
      \hline
      Q_1 - Q_2 & -1 & -1 & 0 & -1 & 0 & -1 & -1 & -1 & 0 & -1 & 0 & -1 \\
      \hline
      Q_R & 0 & 0 & -1 & 0 & 1 & 0 & 0 & 0 & -1 & 0 & 1 & 0 \\
      \hline
      Q_f & 1 & 1 & 1 & 1 & 1 & 1 & -1 & -1 & -1 & -1 & -1 & -1
    \end{array}
  \ee
The components of the chiral and the twisted chiral field strength, $\mathbb{F}$ and $\tilde{\mathbb{F}}$, have the following charge assignments:
\be\label{charge-2}
\begin{array}{c||c|c|c||c|c|c|c}
  {} & \tilde{\phi} & \tilde{\psi}_+ & \tilde{\psi}_- & \sigma & \bar{\lambda}_+ & \lambda_- & A_\mu \\
  \hline
  Q_1 - Q_2 & 1 & 2 & 1 & -1 & 0 & -1 & 0\\
  \hline
  Q_R & 1 & 0 & 1 & 1 & 0 & 1 & 0\\
  \hline
  Q_f & 0 & 0 & 0 & 0 & 0 & 0 & 0
\end{array}
\ee
The fugacities corresponding to $Q_f$, $Q_1 - Q_2$ and $Q_R$ are denoted by $\xi_1$, $\xi_2$ and $z$ respectively.

As we discussed before, the constrained semichiral vector multiplet and the unconstrained semichiral vector multiplet differ by a F-term, which does not show up in the result of localization, hence we can make use of the 1-loop determinant from the previous section. Then for the GLSM given by Eq.~\eqref{eq:semi-EH}, the 1-loop determinant is
\be
  Z_{1-loop}^{EH} = Z_{\tilde{\mathbb{F}}, \mathbb{F}} \cdot Z_1^{L, R} \cdot Z_2^{L, R}\, ,
\ee
where
\begin{align}
  Z_{\tilde{\mathbb{F}}, \mathbb{F}} & = \frac{i \eta(q)^3}{\vartheta_1 (\tau, \xi_2 - z)}\cdot \frac{\vartheta_1 (\tau, 2 \xi_2)}{\vartheta_1 (\tau, \xi_2 +z)}\, ,\nonumber\\
  Z_1^{L, R} & = \frac{\vartheta_1(\tau, u + \xi_1 - z)}{\vartheta_1(\tau, u + \xi_1 - \xi_2)}\cdot \frac{\vartheta_1(\tau, u + \xi_1 + z)}{\vartheta_1(\tau, u + \xi_1 + \xi_2)}\, ,\nonumber\\
  Z_2^{L, R} & = \frac{\vartheta_1(\tau, u - \xi_1 - z)}{\vartheta_1(\tau, u - \xi_1 - \xi_2)}\cdot \frac{\vartheta_1 (\tau, u - \xi_1 + z)}{\vartheta_1 (\tau, u - \xi_1 + \xi_2)}\, .
\end{align}
Then the elliptic genus is given by
\be\label{eq:SemiGenResult}
  Z^{EH} (\tau; z, \xi) = \frac{1}{|W|} \sum_{u_* \in \mathfrak{M}_{\textrm{sing}}^*} \textrm{JK-Res}_{u_*} (Q(u_*), \eta)\, Z_{1-loop} (u)\, ,
\ee
where ``$\textrm{JK-Res}$'' denotes the Jeffrey-Kirwan residue, which was discussed in detail in Refs.~\cite{JK-1, JK-2, Benini-genus-1, Benini-genus-2, Lee} and also briefly reviewed in Appendix~\ref{App:JK}. In practice, the Jeffrey-Kirwan residue can be calculated as follows:
\be
  Z = - \sum_{u_j \in \mathfrak{M}_{\textrm{sing}}^+} \oint_{u=u_j} du\, Z_{1-loop}\, ,
\ee
where we choose $\eta > 0$ for the vector $\eta$ discussed in Appendix~\ref{App:JK}. The poles are at
\be
  Q_i u + \frac{R_i}{2} z + P_i (\xi) = 0 \quad (\textrm{mod}\,\, \mathbb{Z} + \tau \mathbb{Z})\, ,
\ee
where $\xi$ denotes the holonomy of the flavor symmetry on the torus, and $P_i$ are the flavor charges under the maximal torus of the flavor symmetry group $G_F$. The poles with $Q_i > 0$ and $Q_i < 0$ are grouped in to $\mathfrak{M}_{\textrm{sing}}^+$ and $\mathfrak{M}_{\textrm{sing}}^-$ respectively. In the Eguchi-Hanson case, for instance for the phase where the intersection of $H_X = \{u + \xi_1 - \xi_2 = 0 \}$ and $H_Y = \{ u - \xi_1 - \xi_2 = 0 \}$ contributes,
\be
  \mathfrak{M}_{\textrm{sing}}^+ = \{-\xi_1 + \xi_2,\, \xi_1 + \xi_2 \}\, .
\ee
Hence, the elliptic genus equals
\be
  Z^{EH} (\tau; z, \xi) = \frac{\vartheta_1 (\tau, -2 \xi_1 + \xi_2 - z)\cdot \vartheta_1 (\tau, 2 \xi_1 - \xi_2 - z)}{\vartheta_1 (\tau, - 2 \xi_1)\cdot \vartheta(\tau, 2 \xi_1 - 2 \xi_2)} + \frac{\vartheta_1 (\tau, 2 \xi_1 + \xi_2 - z) \cdot \vartheta_1 (\tau, -2 \xi_1 - \xi_2 - z)}{\vartheta_1 (\tau, 2 \xi_1) \cdot \vartheta_1(\tau, -2 \xi_1 -2 \xi_2)}\, ,
\ee
which is the same as the result obtained in Ref.~\cite{Lee}.

From our construction of the ALE space using semichiral GLSM, it is also clear that the elliptic genus for the ALE space coincides with the one for the six-dimensional conifold space. The reason is following. As we discussed before, to obtain an ALE space through a semichiral GLSM we need the semichiral vector multiplet, which has three real components, while to construct a conifold (or resolved conifold when the FI parameter $t \neq 0$) one should use the constrained semichiral vector multiplet, which has only one real component. However, these two vector multiplets differ only by a superpotential term, which does not affect the result of the localization. Hence, the result that we obtained using localization give us the elliptic genus both for the ALE space and for the conifold.\footnote{We would like to thank P. Marcos Crichigno for discussing this.}

\subsection{Taub-NUT space}

Taub-NUT space is an example of the ALF space, and can be constructed by semichiral GLSM as follows~\cite{Marcos}:
\begin{align}
  {\cal L} & = \int d^4 \theta \Big[-\frac{1}{2 e^2} (\bar{\tilde{\mathbb{F}}} \tilde{\mathbb{F}} - \bar{\mathbb{F}} \mathbb{F}) + \bar{\mathbb{X}}_1^L \, e^{V_L} \, \mathbb{X}_1^L + \bar{\mathbb{X}}_1^R \, e^{V_R} \, \mathbb{X}_1^R + \alpha (\bar{\mathbb{X}}_1^L \, e^{i \bar{\tilde{\mathbb{V}}}} \, \mathbb{X}_1^R + \bar{\mathbb{X}}_1^R \, e^{-i \tilde{\mathbb{V}}} \, \mathbb{X}_1^L) \nonumber\\
  {} & \quad \quad + \frac{1}{2} \left(\mathbb{X}_2^L + \bar{\mathbb{X}}_2^L + V_L  \right)^2 + \frac{1}{2} \left(\mathbb{X}_2^R + \bar{\mathbb{X}}_2^R + V_R \right)^2 + \frac{\alpha}{2} \left(\mathbb{X}_2^L + \bar{\mathbb{X}}_2^R - i \tilde{\mathbb{V}} \right)^2 + \frac{\alpha}{2} \left(\mathbb{X}_2^R + \mathbb{X}_2^L + i \bar{\tilde{\mathbb{V}}} \right)^2 \Big] \nonumber\\
  {} & \quad \quad + \left( \int d^2 \theta \, \Phi\, \mathbb{F} + c.c. \right) - \left( \int d^2 \tilde{\theta} \, t\, \tilde{\mathbb{F}} + c.c. \right)\, ,\label{eq:semi-TN}
\end{align}
where for simplicity we set $t = 0$.

Using the results from the previous section, and assigning the same R-symmetry and the flavor symmetry charges as in the Eguchi-Hanson case \eqref{charge-1} \eqref{charge-2}, we can write down immediately the 1-loop contribution from the semichiral vector multiplet, $\tilde{\mathbb{F}}$ and $\mathbb{F}$, as well as the one from the semichiral multiplet, $\mathbb{X}_1^L$ and $\mathbb{X}_1^R$, of the model \eqref{eq:semi-TN}:
\begin{align}
  Z_{\tilde{\mathbb{F}}, \mathbb{F}} & = \prod_{m, n \in \mathbb{Z}} \frac{n + \tau m - 2 \xi_2}{(n + \tau m - \xi_2 + z) \cdot (n + \tau m - \xi_2 - z)} \cdot \prod_{(m, n) \neq (0, 0)} (n + m \tau)\, ,\label{eq:Z-TN-FF}\\
  Z_1^{L, R} & = \frac{\vartheta_1(\tau, u + \xi_1 - z)}{\vartheta_1(\tau, u + \xi_1 - \xi_2)}\cdot \frac{\vartheta_1(\tau, u + \xi_1 + z)}{\vartheta_1(\tau, u + \xi_1 + \xi_2)}\, .\label{eq:Z-TN-semi}
\end{align}
However, to obtain the full 1-loop determinant, we still have to work out the part of the model from semichiral St\"uckelberg fields, and localize it to obtain its contribution to the 1-loop determinant. Let us start with the Lagrangian for the St\"uckelberg field in the superspace:
\begin{align}\label{eq:Lst}
  {\cal L}_{St} & = \int d^4 \theta \Big[\frac{1}{2} \left(\mathbb{X}_2^L + \bar{\mathbb{X}}_2^L + V_L  \right)^2 + \frac{1}{2} \left(\mathbb{X}_2^R + \bar{\mathbb{X}}_2^R + V_R \right)^2 \nonumber\\
  {} & \qquad\qquad + \frac{\alpha}{2} \left(\mathbb{X}_2^L + \bar{\mathbb{X}}_2^R - i \tilde{\mathbb{V}} \right)^2 + \frac{\alpha}{2} \left(\mathbb{X}_2^R + \mathbb{X}_2^L + i \bar{\tilde{\mathbb{V}}} \right)^2 \Big]\, .
\end{align}
Expanding the Lagrangian into components and integrate out auxiliary fields (see Appendix~\ref{App:St}), we obtain
\begin{align}
  {\cal L}_{St} &= \frac{\alpha - 1}{\alpha} \, (\bar{r}_1 \square r_1 + \bar{\gamma}_1 \square \gamma_1) + \frac{\alpha + 1}{\alpha} \, (\bar{r}_2 \square r_2 + \bar{\gamma}_2 \square \gamma_2) \nonumber\\~
  {} & \quad + \frac{i}{2} (\frac{1}{\alpha^2} - \alpha^2) \bar{\psi}_+^2 D_- \psi_+^2 - \frac{i}{2} (\frac{1}{\alpha^2} - \alpha^2) \bar{\psi}_-^2 D_+ \psi_-^2 + \bar{\chi}_-^L 2 i D_+ \chi_-^L - \bar{\chi}_+^R 2 i D_- \chi_+^R\, .
\end{align}
As discussed in Appendix~\ref{App:St}, among the real components $r_{1, 2}$ and $\gamma_{1, 2}$ only $r_2$ transforms under the gauge transformations. We can assign the following charges to the components of the St\"uckelberg field:
\be\label{charge-3}
\begin{array}{c||c|c|c|c|c|c|c|c}
  {} & r_1 & r_2 & \gamma_1 & \gamma_2 & \psi_+ & \psi_- & \chi_+ & \chi_-\\
  \hline
  Q_1 - Q_2 & -2 & 0 & -2 & 0 & -1 & -2 & -1 & 0\\
  \hline
  Q_R & 0 & 0 & 0 & 0 & -1 & 0 & 1 & 0\\
  \hline
  Q_f & 0 & 0 & 0 & 0 & 0 & 0 & 0 & 0
\end{array}
\ee
Taking both the momentum and the winding modes into account, we obtain the contribution from the St\"uckelberg field to the 1-loop determinant
\be
  Z_{St} = \prod_{m, n\in \mathbb{Z}} \frac{(n + \tau m + \xi_2 + z) \cdot (n + \tau m + \xi_2 - z)}{n + \tau m + 2 \xi_2} \cdot \prod_{(m, n) \neq (0, 0)} \frac{1}{n + m\tau} \cdot \sum_{v, w \in \mathbb{Z}} e^{-\frac{g^2 \pi}{\tau_2} |u + v + \tau w|^2}\, .
\ee
Together with Eq.~\eqref{eq:Z-TN-FF} and Eq.~\eqref{eq:Z-TN-semi}, we obtain the full 1-loop determinant of the elliptic genus for the Taub-NUT space
\be
  Z_{1-loop}^{TN} = \frac{\vartheta_1(\tau, u + \xi_1 - z)}{\vartheta_1(\tau, u + \xi_1 - \xi_2)}\cdot \frac{\vartheta_1(\tau, u + \xi_1 + z)}{\vartheta_1(\tau, u + \xi_1 + \xi_2)} \cdot \sum_{v, w \in \mathbb{Z}} e^{-\frac{g^2 \pi}{\tau_2} |u + v + \tau w|^2}\, .
\ee
The elliptic genus for the Taub-NUT space is given by
\be\label{eq:TNgenus}
  Z^{TN} = g^2 \int_{E(\tau)} \frac{du\, d\bar{u}}{\tau_2}\, \frac{\vartheta_1(\tau, u + \xi_1 - z)}{\vartheta_1(\tau, u + \xi_1 - \xi_2)}\cdot \frac{\vartheta_1(\tau, u + \xi_1 + z)}{\vartheta_1(\tau, u + \xi_1 + \xi_2)} \cdot \sum_{v, w \in \mathbb{Z}} e^{-\frac{g^2 \pi}{\tau_2} |u + v + \tau w|^2}\, ,
\ee
where $E(\tau) = \mathbb{C} / (\mathbb{Z} + \tau \mathbb{Z})$. This result is the same as the one in Ref.~\cite{Lee} obtained from the chiral GLSM. We would like to emphasize that the result Eq.~\eqref{eq:TNgenus} cannot be included in the result presented in Eq.~\eqref{eq:SemiGenResult}. The reason is that one needs the semichiral St\"uckelberg field to describe the Taub-NUT space, which has the holomorphic anomaly in the elliptic genus. Some related discussions on the holomorphic anomaly in the elliptic genus of the noncompact space can be found in Refs.~\cite{Murthy, Lee}.

Similar to the ALE space, the elliptic genus for the ALF space should coincide with the one for some six-dimensional space. In semichiral GLSM language, one is obtained using the unconstrained semichiral vector multiplet, while the other is constructed using the constrained semichiral vector multiplet. However, as far as we know, this kind of six-dimensional space is not well studied in the literature as the conifold. We would like to investigate it in more detail in the future.

\section{Low energy effective twisted superpotential}\label{sec:LEETS}

In this section, we attempt to study the low energy physics of a general GLSM.

\be
    \frac{1}{2\pi} \int \Delta \mathcal{L}_E^{(1)} d^2 x = \textrm{log}\, \textrm{det}\, \Delta_{\textrm{bos}} - \textrm{log}\, \textrm{det}\, \Delta_{\textrm{ferm}}\, ,
  \ee
  where
  \be
    \Delta_{\textrm{bos}} \equiv
     \left(\begin{array}{cc}
      \square + D + \alpha^2 |\hat{\sigma}|^2 & \frac{1}{\alpha} \square + \alpha D + \alpha |\hat{\sigma}|^2\\
      \frac{1}{\alpha} \square + \alpha D + \alpha |\hat{\sigma}|^2 & \square + D + \alpha^2 |\hat{\sigma}|^2
    \end{array} \right)
  \ee
  is the $2\times 2$-matrix appearing in $\mathcal{L}_{\textrm{bos}}$, while $\Delta_{\textrm{ferm}}$ is the corresponding matrix from the fermionic sector, which is irrelevant for the calculation of the 1-loop coupled to the field $D$. Up to an irrelevant constant due to field rescaling, we obtain
  \begin{align}
    \textrm{log}\, \textrm{det}\, \Delta_{\textrm{bos}} & = \textrm{log} \left(\alpha D + D_\mu D^\mu + \alpha^2 \sigma^2 \right) + \textrm{log} \left(-\alpha D + D_\mu D^\mu + \alpha^2 \sigma^2 \right)\nonumber\\
    {} & = \frac{\alpha D}{D_\mu D^\mu + \alpha^2 \sigma^2} + \frac{-\alpha D}{D_\mu D^\mu + \alpha^2 \sigma^2} + (\textrm{higher-order terms in } D)\, .\label{eq:logdetDelta}
  \end{align}
  Since the terms linear in $D$ have opposite signs, they cancel out exactly and do not show up in $\Delta \mathcal{L}_E^{(1)}$. Therefore, the effective twisted superpotential $\widetilde{W}$ is zero, and do not have terms like $\Sigma\, \textrm{log} (\Sigma)$.

As discussed before, we can turn on two types of twisted masses $m_i$ and $\widetilde{m}_i$ using the semichiral vector multiplet. However, $m_i$ can be viewed as a chiral superfield and cannot enter the twisted superpotential. On the other hand, the effect of $\widetilde{m}_i$ is merely a shift of $\hat{\sigma}$
\begin{displaymath}
  \hat{\sigma} \rightarrow \hat{\sigma} + \widetilde{m}_i\, .
\end{displaymath}
Although the shifts $\widetilde{m}_i$ are generically different for different flavors, the nontrivial contributions to the effective twisted superpotential cancel out within each flavor, as one can see from Eq.~\eqref{eq:logdetDelta}. Hence, similar to the case without twisted masses, after turning on the twisted masses the effective twisted superpotential still remains zero.

We may ask whether there is a better way to understand why the effective twisted superpotential vanishes. Indeed, a more conceptual reasoning goes as follows. We notice that the theory without twisted mass deformations is invariant under a larger supersymmetry, namely it has $\CN=(4,4)$ rather than just $\CN=(2,2)$ supersymmetry \cite{Goteman}. Since a nontrivial twisted superpotential is not compatible with $\CN=(4,4)$ supersymmetry, we cannot generate an effective twisted superpotential term at the low energy. This fact is not completely new. If we construct a GLSM using chiral superfields in such a way that the target space is a hyperk\"ahler manifold, the contributions from chiral superfields will cancel in pairs and the final result is zero. However, in a GLSM with chiral superfields, after we turning on twisted masses, the $\CN=(4,4)$ supersymmetry algebra is broken down to its $\CN=(2,2)$ subalgebra, and we obtain a nonzero effective twisted superpotential. Hence the real question is why the twisted mass deformations for semichiral superfields do not break $\CN=(4,4)$ supersymmetry. To answer this question, we take a slightly different point of view of the twisted masses. Instead of setting component fields to the required constant values by hand, we introduce Lagrange multipliers $\hat{\Sigma}$ and $\hat{\widetilde{\Sigma}}$, \footnote{We would like to thank Martin Ro\v cek for discussing this.}
\be
  \int d^2 \theta\, \hat{\Sigma} (\mathbb{F} - m) + \int d^2 \tilde{\theta}\, \hat{\widetilde{\Sigma}} (\widetilde{\mathbb{F}} - \widetilde{m})\, ,
\ee
where $\hat{\Sigma}$ is a chiral superfield, while $\hat{\widetilde{\Sigma}}$ is a twisted chiral superfield. Since now we focus on the flavor symmetry group, let us suppress the color indices at the moment, then the K\"ahler potential part of the Lagrangian is
\be
  \int d^4 \theta\, \left[\bar{\mathbb{X}}_L e^{V_L} \mathbb{X}_L + \bar{\mathbb{X}}_R e^{V_R} \mathbb{X}_R + \alpha (\bar{\mathbb{X}}_L e^{i \bar{\widetilde{\mathbb{V}}}} \mathbb{X}_R + \bar{\mathbb{X}}_R e^{-i \widetilde{\mathbb{V}}} \mathbb{X}_L)\right]\, ,
\ee
or written in covariant approach as
\be
  \int d^4 \theta\, \left[\bar{\mathbb{X}}_L \mathbb{X}_L + \bar{\mathbb{X}}_R \mathbb{X}_R + \alpha (\bar{\mathbb{X}}_L \mathbb{X}_R + \bar{\mathbb{X}}_R \mathbb{X}_L)\right]\, ,
\ee
where now
\be
  \mathbb{F} = \{\bar{\mathbb{D}}_+ , \, \bar{\mathbb{D}}_-\}\, ,\quad \tilde{\mathbb{F}} = \{\bar{\mathbb{D}}_+ , \, \mathbb{D}_-\}\, .
\ee
Next, we need to work out the expression of the K\"ahler potential part of the Lagrangian in components, and integrating out the Lagrange multipliers $\hat{\Sigma}$ and $\hat{\widetilde{\Sigma}}$ will set the lowest components $\mathbb{F}$ and $\widetilde{\mathbb{F}}$ to some constant values $m$ and $\widetilde{m}$ respectively, which are the twisted masses in our model. To be precise, the superfields carry flavor indices as follows:
\be
  \int d^4 \theta\, \left[\bar{\mathbb{X}}_L^i \mathbb{X}_{Li} + \bar{\mathbb{X}}_R^i \mathbb{X}_{Ri} + \alpha (\bar{\mathbb{X}}_L^i \mathbb{X}_{Ri} + \bar{\mathbb{X}}_R^i \mathbb{X}_{Li})\right] + \int d^2 \theta\, \hat{\Sigma}^{ij} (\mathbb{F}_{ij} - m_{ij}) + \int d^2 \tilde{\theta}\, \hat{\widetilde{\Sigma}}^{ij} (\widetilde{\mathbb{F}}_{ij} - \widetilde{m}_{ij})\, .
\ee
However, the terms
\begin{displaymath}
  \int d^2 \theta\, \hat{\Sigma}^{ij} m_{ij}\quad \textrm{and}\quad \int d^2 \tilde{\theta}\, \hat{\widetilde{\Sigma}}^{ij} \widetilde{m}_{ij}
\end{displaymath}
are similar to the FI term, which is gauge invariant only for Abelian groups. Hence, only diagonal parts of the matrices $m_{ii}$ and $\tilde{m}_{ii}$ preserve the gauge symmetry, and we only need to consider
\be
  \int d^2 \theta\, \hat{\Sigma}^i (\mathbb{F}_i - m_i) + \int d^2 \tilde{\theta}\, \hat{\widetilde{\Sigma}}^i (\widetilde{\mathbb{F}}_i - \widetilde{m}_i)\, .\label{eq:twMassSVM}
\ee
$m_i \neq 0$ will break the R-symmetry. Nevertheless, since $m_i$ is part of the superpotential, it does not enter the effective twisted superpotential in the end. On the other hand, $\widetilde{m}_i$ plays the same role as Coulomb branch moduli $\hat{\sigma}$, which is essentially the VEV of the scalar in the vector multiplet. Since using the semichiral vector multiplets to gauge semichiral multiplets preserves $\mathcal{N}=(4,4)$ supersymmetry, it is impossible to generate a nonzero twisted superpotential term in this way.

\section{Conclusion and future directions}\label{sec:con}

In this paper, we have studied the dynamics of GLSMs with semichiral superfields on the flat space.

We have computed the elliptic genus using both the Hamiltonian formalism and the path integral formalism. We have also worked out two important examples, namely the Eguchi-Hanson space and the Taub-NUT space. The results agree with the previous computations using GLSMs without using semichiral superfields~\cite{Lee}.

It is natural to ask whether our computation can be generalized to other models. There are many interesting cases which do not have known realization in terms of GLSMs. For example, we may construct Wess-Zumino-Witten models with manifest $\CN = (2, 2)$ supersymmetry. The elliptic genus can be again computed with a minor modification. It will be described in detail in the subsequent paper \cite{followup}.

Futhermore, we have also computed the low energy effective twisted superpotential $\tilde{\CW}^{\rm eff}$ of the GLSMs on the Coulomb branch. Unfortunately, the contribution from semichiral superfields to $\tilde{\CW}^{\rm eff}$ vanishes. Therefore, the low energy behavior of the GLSM with semichiral superfields is determined only by the generalized K\"ahler potential, which is not protected by supersymmetry and is difficult to compute exactly. It will be interesting if one can figure out some other methods to describe some exact properties of the low energy effective theory.

\section*{Acknowledgments}
We would like to thank Francesco Benini, Marcos Crichigno, Abhijit Gadde, Sergei Gukov, Dharmesh Jain, Sungjay Lee, Nikita Nekrasov, Daniel Park, Yachao Qian, Martin Ro\v cek, Peter van Nieuwenhuizen and Peng Zhao for discussions. The research of J. N. is supported in part by NSF grant No. PHY-1316617. The work of X. Z. is supported in part by NSF grant No. PHY-1404446.

\newpage
\appendix
\section{Two-dimensional $\CN=(2,2)$ superspace}\label{App:superspace}

The bosonic coordinates of the superspace are $x^{\mu}, \mu=0,1$. We take the flat Minkowski metric to be $\eta_{\mu\nu}={\rm diag} (-1, 1)$. The fermionic coordinates of the superspace are $\theta^+$, $\theta^-$, $\bar{\theta}^+$ and $\bar{\theta}^-$, with the complex conjugation relation $(\theta^\pm)^* = \bar{\theta}^\pm$. The indices $\pm$ stand for the chirality under a Lorentz transformation. To raise or lower the spinor index, we use
\be
    \psi_\alpha = \epsilon_{\alpha\beta}\, \psi^\beta\, ,\quad \psi^\alpha = \epsilon^{\alpha\beta}\, \psi_\beta\, ,
\ee
where
\be
    \epsilon_{\alpha\beta} = \left(
    \begin{array}{cc}
      0 & -1 \\
      1 & 0
    \end{array} \right)\, ,\quad
    \epsilon^{\alpha\beta} = \left(
    \begin{array}{cc}
      0 & 1 \\
      -1 & 0
    \end{array} \right)\, ,\quad
    \alpha, \beta = - , +\, .
  \ee
Hence, we have $\psi_+ = \psi^-$, $\psi_- = -\psi^+$.

The supercharges and the supercovariant derivative operators are
\ben
    Q_\pm &=& \frac{\partial}{\partial \theta^\pm} + i \bar{\theta}^\pm \partial_\pm\, ,\quad \bar{Q}_\pm = - \frac{\partial}{\partial \bar{\theta}^\pm} - i \theta^\pm \partial_\pm\, , \\~
    \mathbb{D}_\pm &=& \frac{\partial}{\partial \theta^\pm} - i \bar{\theta}^\pm \partial_\pm\, ,\quad \bar{\mathbb{D}}_\pm = - \frac{\partial}{\partial \bar{\theta}^\pm} + i \theta^\pm \partial_\pm\, ,
\een
where
\be
    \partial_\pm \equiv \frac{1}{2} \left(\frac{\partial}{\partial x^0} \pm \frac{\partial}{\partial x^1} \right)\, .
\ee
They satisfy the anti-commutation relations
\be
    \{Q_\pm,\, \bar{Q}_\pm \} = -2 i \partial_\pm\, ,\quad \{\mathbb{D}_\pm,\, \bar{\mathbb{D}}_\pm \} = 2 i \partial_\pm\, ,
\ee
with all the other anti-commutators vanishing. In particular,
\be
\{Q_\pm,\, \mathbb{D}_\pm \} = 0 \, .
\ee

\section{Gauged linear sigma model with semichiral superfields in components}\label{App:GLSMsemi}

If we expand the theory~(\ref{GLSMsemi}) in components, we obtain the Lagrangian
  \begin{align}
    \mathcal{L}_{SC} & = \bar{X}^L 2 i D_- 2 i D_+ X^L - \bar{X}^L (\sigma_1^2 + \sigma_2^2) X^L + \bar{X}^L D X^L + \bar{F}^L F^L \nonumber\\
    {} & \quad - \bar{M}_{-+}^L M_{-+}^L - \bar{M}_{--}^L 2 i D_+ X^L - \bar{X}^L 2 i D_+ M_{--}^L + \bar{M}_{-+}^L \bar{\hat{\sigma}} X^L + \bar{X}^L \hat{\sigma} M_{-+}^L \nonumber\\
    {} & \quad + \bar{\psi}_-^L 2 i D_+ \psi_-^L + \bar{\psi}_+^L 2 i D_- \psi_+^L - \bar{\psi}_-^L \hat{\sigma} \psi_+^L - \bar{\psi}_+^L \bar{\hat{\sigma}} \psi_-^L \nonumber\\
    {} & \quad + \bar{X}^L i \lambda_+ \psi_-^L - \bar{X}^L i \lambda_- \psi_+^L + \bar{\psi}_+^L i \bar{\lambda}_- X^L - \bar{\psi}_-^L i \bar{\lambda}_+ X^L - \bar{\eta}_-^L \psi_+^L - \bar{\psi}_+^L \eta_-^L \nonumber\\
    {} & \quad - \bar{\chi}_-^L 2 i D_+ \chi_-^L + \bar{X}^L i \bar{\lambda}_+ \chi_-^L - \bar{\chi}_-^L i \lambda_+ X^L \nonumber\\
    {} & \quad  + \bar{X}^R 2 i D_- 2 i D_+ X^R - \bar{X}^R (\sigma_1^2 + \sigma_2^2) X^R + \bar{X}^R D X^R + \bar{F}^R F^R \nonumber\\
    {} & \quad - \bar{M}_{+-}^R M_{+-}^R - \bar{M}_{++} 2 i D_- X^R - \bar{X}^R 2 i D_- M_{++}^R + \bar{M}_{+-}^R \hat{\sigma} X^R + \bar{X}^R \bar{\hat{\sigma}} M_{+-}^R \nonumber\\
    {} & \quad + \bar{\psi}_-^R 2 i D_+ \psi_-^R + \bar{\psi}_+^R 2 i D_- \psi_+^R - \bar{\psi}_-^R \hat{\sigma} \psi_+^R - \bar{\psi}_+^R \bar{\hat{\sigma}} \psi_-^R \nonumber\\
    {} & \quad + \bar{X}^R i \lambda_+ \psi_-^R - \bar{X}^R i \lambda_- \psi_+^R + \bar{\psi}_+^R i \bar{\lambda}_- X^R - \bar{\psi}_-^R i \bar{\lambda}_+ X^R + \bar{\eta}_+^R \psi_-^R + \bar{\psi}_-^R \eta_+^R \nonumber\\
    {} & \quad - \bar{\chi}_+^R 2 i D_- \chi_+^R - \bar{X}^R i \bar{\lambda}_- \chi_+^R + \bar{\chi}_+^R i \lambda_- X^R \nonumber\\
    {} & \quad + \alpha \bar{X}^L 2 i D_- 2 i D_+ X^R - \alpha \bar{X}^L (\sigma_1^2 + \sigma_2^2) X^R + \alpha  \bar{X}^L D X^R + \alpha \bar{F}^L F^R \nonumber\\
    {} & \quad + \alpha \bar{M}_{--}^L M_{++}^R - \alpha \bar{M}_{--}^L 2 i D_+ X^R - \alpha \bar{X}^L 2 i D_- M_{++}^R + \alpha \bar{M}_{-+}^L \bar{\hat{\sigma}} X^R + \alpha \bar{X}^L \bar{\hat{\sigma}} M_{+-}^R \nonumber\\
    {} & \quad + \alpha \bar{\psi}_-^L 2 i D_+ \psi_-^R + \alpha \bar{\psi}_+^L 2 i D_- \psi_+^R - \alpha \bar{\psi}_-^L \hat{\sigma} \psi_+^R - \alpha \bar{\psi}_+^L \bar{\hat{\sigma}} \psi_-^R \nonumber\\
    {} & \quad + \alpha \bar{X}^L i \lambda_+ \psi_-^R - \alpha \bar{X}^L i \lambda_- \psi_+^R + \alpha \bar{\psi}_+^L i \bar{\lambda}_- X^R - \alpha \bar{\psi}_-^L i \bar{\lambda}_+ X^R - \alpha \bar{\eta}_-^L \psi_+^R + \alpha \bar{\psi}_-^L \eta_+^R \nonumber\\
    {} & \quad + \alpha \bar{\chi}_-^L \bar{\hat{\sigma}} \chi_+^R - \alpha \bar{X}^L i \bar{\lambda}_- \chi_+^R - \alpha \bar{\chi}_-^L i \lambda_+ X^R \nonumber\\
    {} & \quad + \alpha \bar{X}^R 2 i D_- 2 i D_+ X^L - \alpha \bar{X}^R (\sigma_1^2 + \sigma_2^2) X^L + \alpha  \bar{X}^R D X^L + \alpha \bar{F}^R F^L \nonumber\\
    {} & \quad + \alpha \bar{M}_{++}^R M_{--}^L - \alpha \bar{M}_{++}^R 2 i D_- X^L - \alpha \bar{X}^R 2 i D_+ M_{--}^L + \alpha \bar{M}_{+-}^R \hat{\sigma} X^L + \alpha \bar{X}^R \hat{\sigma} M_{-+}^L \nonumber\\
    {} & \quad + \alpha \bar{\psi}_-^R 2 i D_+ \psi_-^L + \alpha \bar{\psi}_+^R 2 i D_- \psi_+^L - \alpha \bar{\psi}_-^R \hat{\sigma} \psi_+^L - \alpha \bar{\psi}_+^R \bar{\hat{\sigma}} \psi_-^L \nonumber\\
    {} & \quad + \alpha \bar{X}^R i \lambda_+ \psi_-^L - \alpha \bar{X}^R i \lambda_- \psi_+^L + \alpha \bar{\psi}_+^R i \bar{\lambda}_- X^L - \alpha \bar{\psi}_-^R i \bar{\lambda}_+ X^L + \alpha \bar{\eta}_+^R \psi_-^L - \alpha \bar{\psi}_+^R \eta_-^L \nonumber\\
    {} & \quad + \alpha \bar{\chi}_+^R \hat{\sigma} \chi_-^L + \alpha \bar{X}^R i \bar{\lambda}_+ \chi_-^L + \alpha \bar{\chi}_+^R i \lambda_- X^L\, .\label{eq:theoryMink}
  \end{align}

  The supersymmetry transformation laws for the abelian vector multiplet are
  \begin{align}
    \delta A_\mu & = \frac{i}{2} \epsilon \sigma_\mu \bar{\lambda} + \frac{i}{2} \bar{\epsilon} \sigma_\mu \lambda\, ,\nonumber\\
    \delta \hat{\sigma} & = - i \epsilon_- \bar{\lambda}_+ - i \bar{\epsilon}_+ \lambda_- \, ,\nonumber\\
    \delta \bar{\hat{\sigma}} & = - i \epsilon_+ \bar{\lambda}_- - i \bar{\epsilon}_- \lambda_+\, ,\nonumber\\
    \delta \lambda_+ & = 2 \epsilon_- \partial_+ \bar{\hat{\sigma}} + i \epsilon_+ D - \epsilon_+ F_{01}\, ,\nonumber\\
    \delta \lambda_- & = 2 \epsilon_+ \partial_- \hat{\sigma} + i \epsilon_- D + \epsilon_- F_{01}\, ,\nonumber\\
    \delta \bar{\lambda}_+ & = 2 \bar{\epsilon}_- \partial_+ \hat{\sigma} - i \bar{\epsilon}_+ D - \bar{\epsilon}_+ F_{01}\, ,\nonumber\\
    \delta \bar{\lambda}_- & = 2 \bar{\epsilon}_+ \partial_- \bar{\hat{\sigma}} - i \bar{\epsilon}_- D + \bar{\epsilon}_- F_{01}\, ,\nonumber\\
    \delta D & = \epsilon_+ \partial_- \bar{\lambda}_+ + \epsilon_- \partial_+ \bar{\lambda}_- - \bar{\epsilon}_+ \partial_- \lambda_+ - \bar{\epsilon}_- \partial_+ \bar{\lambda}_-\, ,
  \end{align}
  where $F_{01}=\partial_0 A_1 - \partial_1 A_0$, and
  \be
    \sigma_0 =
    \left( \begin{array}{cc}
      1 & 0 \\
      0 & 1
    \end{array} \right)\, ,\quad
    \sigma_1 =
    \left( \begin{array}{cc}
      1 & 0\\
      0 & -1
    \end{array} \right)\, .
  \ee

The supersymmetry transformations for the components of semichiral multiplets $\mathbb{X}$ are
  \begin{align}
    \delta X & = \epsilon \psi + \bar{\epsilon} \chi \, , \nonumber\\
    \delta \psi_+ & = - \epsilon_+ F - \bar{\epsilon}_+ \bar{\hat{\sigma}} X + \bar{\epsilon}_+ M_{-+} + \bar{\epsilon}_- 2 i D_+ X - \bar{\epsilon}_- M_{++}\, ,\nonumber\\
    \delta \psi_- & = - \epsilon_- F - \bar{\epsilon}_+ 2 i D_- X + \bar{\epsilon}_+ M_{--} + \bar{\epsilon}_- \hat{\sigma} X - \bar{\epsilon}_- M_{+-}\, ,\nonumber\\
    \delta F & = \bar{\epsilon}_+ 2 i D_- \psi_+ + \bar{\epsilon}_- 2 i D_+ \psi_- - \bar{\epsilon}_+ \eta_- + \bar{\epsilon}_- \eta_+ - \bar{\epsilon}_+ \bar{\hat{\sigma}} \psi_- - \bar{\epsilon}_- \hat{\sigma} \psi_+ + i \bar{\epsilon}_+ \bar{\lambda}_- X - i \bar{\epsilon}_- \bar{\lambda}_+ X\, ,\nonumber\\
    \delta \chi_+ & = - \epsilon_- M_{++} + \epsilon_+ M_{+-}\, ,\nonumber\\
    \delta \chi_- & = - \epsilon_- M_{-+} + \epsilon_+ M_{--}\, ,\nonumber\\
    \delta M_{+-} & = - \epsilon_- \eta_+ + \bar{\epsilon}_- \hat{\sigma} \chi_+ - \bar{\epsilon}_+ 2 i D_- \chi_+\, ,\nonumber\\
    \delta M_{-+} & = - \epsilon_+ \eta_- + \bar{\epsilon}_- 2 i D_+ \chi_- - \bar{\epsilon}_+ \bar{\hat{\sigma}} \chi_-\, ,\nonumber\\
    \delta M_{++} & = -\epsilon_+ \eta_+ + \bar{\epsilon}_- 2 i D_+ \chi_+ - \bar{\epsilon}_+ \bar{\hat{\sigma}} \chi_+\, ,\nonumber\\
    \delta M_{--} & = -\epsilon_- \eta_- + \bar{\epsilon}_- \hat{\sigma} \chi_- - \bar{\epsilon}_+ 2 i D_- \chi_-\, ,\nonumber\\
    \delta \eta_+ & = \bar{\epsilon}_- 2 i D_+ M_{+-} - \bar{\epsilon}_- i \bar{\lambda}_+ \chi_+ - \bar{\epsilon}_- \hat{\sigma} M_{++} - \bar{\epsilon}_+ \bar{\hat{\sigma}} M_{+-} + \bar{\epsilon}_+ i \bar{\lambda}_- \chi_+ + \bar{\epsilon}_+ 2 i D_- M_{++}\, ,\nonumber\\
    \delta \eta_- & = \bar{\epsilon}_- 2 i D_+ M_{--} - \bar{\epsilon}_- i \bar{\lambda}_+ \chi_- - \bar{\epsilon}_- \hat{\sigma} M_{-+} - \bar{\epsilon}_+ \bar{\hat{\sigma}} M_{--} + \bar{\epsilon}_+ i \bar{\lambda}_- \chi_- + \bar{\epsilon}_+ 2 i D_- M_{-+}\, ,
  \end{align}
   and similarly for $\bar{\mathbb{X}}$
  \begin{align}
    \delta \bar{X} & = \epsilon \bar{\chi} + \bar{\epsilon} \bar{\psi}\, ,\nonumber\\
    \delta \bar{\psi}_+ & = \epsilon_- 2 i D_+ \bar{X} + \epsilon_- \bar{M}_{++} - \epsilon_+ \hat{\sigma} \bar{X} - \epsilon_+ \bar{M}_{-+} + \bar{\epsilon}_+ \bar{F}\, ,\nonumber\\
    \delta \bar{\psi}_- & = \epsilon_- \bar{\hat{\sigma}} \bar{X} + \epsilon_- \bar{M}_{+-} - \epsilon_+ 2 i D_- \bar{X} - \epsilon_+ \bar{M}_{--} + \bar{\epsilon}_- \bar{F}\, ,\nonumber\\
    \delta \bar{F} & = \epsilon_+ 2 i D_- \bar{\psi}_+ + \epsilon_- 2 i D_+ \bar{\psi}_- + \epsilon_+ \bar{\eta}_- - \epsilon_- \bar{\eta}_+ - \epsilon_+ \hat{\sigma} \bar{\psi}_- - \epsilon_- \bar{\hat{\sigma}} \bar{\psi}_+ - \epsilon_+ i \lambda_- \bar{X} + \epsilon_- i \lambda_+ \bar{X}\, ,\nonumber\\
    \delta \bar{\chi}_+ & = \bar{\epsilon}_- \bar{M}_{++} - \bar{\epsilon}_+ \bar{M}_{+-}\, ,\nonumber\\
    \delta \bar{\chi}_- & = \bar{\epsilon}_- \bar{M}_{-+} - \bar{\epsilon}_+ \bar{M}_{--}\, ,\nonumber\\
    \delta \bar{M}_{+-} & = \epsilon_- \bar{\hat{\sigma}} \bar{\chi}_+ - \epsilon_+ 2 i D_- \bar{\chi}_+ + \bar{\epsilon}_- \bar{\eta}_+\, ,\nonumber\\
    \delta \bar{M}_{-+} & = \epsilon_- 2 i D_+ \bar{\chi}_- - \epsilon_+ \hat{\sigma} \bar{\chi}_- + \bar{\epsilon}_+ \bar{\eta}_-\, ,\nonumber\\
    \delta \bar{M}_{++} & = \epsilon_- 2 i D_+ \bar{\chi}_+ - \epsilon_+ \hat{\sigma} \bar{\chi}_+ + \bar{\epsilon}_+ \bar{\eta}_+\, ,\nonumber\\
    \delta \bar{M}_{--} & = \epsilon_- \bar{\hat{\sigma}} \bar{\chi}_- - \epsilon_+ 2 i D_- \bar{\chi}_- + \bar{\epsilon}_- \bar{\eta}_-\, ,\nonumber\\
    \delta \bar{\eta}_+ & = \epsilon_- 2 i D_+ \bar{M}_{+-} - \epsilon_- i \bar{\lambda}_+ \bar{\chi}_+ - \epsilon_- \bar{\hat{\sigma}} \bar{M}_{++} - \epsilon_+ \hat{\sigma} \bar{M}_{+-} + \epsilon_+ i \bar{\lambda}_- \bar{\chi}_+ + \epsilon_+ 2 i D_- \bar{M}_{++}\, ,\nonumber\\
    \delta \bar{\eta}_- & = \epsilon_- 2 i D_+ \bar{M}_{--} - \epsilon_- i \bar{\lambda}_+ \bar{\chi}_- - \epsilon_- \bar{\hat{\sigma}} \bar{M}_{-+} - \epsilon_+ \hat{\sigma} \bar{M}_{--} + \epsilon_+ i \bar{\lambda}_- \bar{\chi}_- + \epsilon_+ 2 i D_- \bar{M}_{-+}\, .
  \end{align}
  The transformation laws are written in the general form, and one should set some fields to be zero after imposing the constraints.

  Varying the fields $M_{--}^L$, $M_{-+}^L$, $M_{++}^R$, $M_{+-}^R$, $\bar{M}_{--}^L$, $\bar{M}_{-+}^L$, $\bar{M}_{++}^R$ and $\bar{M}_{+-}^R$, we obtain
  \begin{align}
    0 & = \alpha \bar{M}_{++}^R + 2 i D_+ \bar{X}^L + \alpha 2 i D_+ \bar{X}^R \, ,\\
    0 & = - \bar{M}_{-+}^L + \bar{X}^L \hat{\sigma} + \alpha \bar{X}^R \hat{\sigma} \, ,\\
    0 & = \alpha \bar{M}_{--}^L + 2 i D_- \bar{X}^R + \alpha 2 i D_- \bar{X}^L \, ,\\
    0 & = - \bar{M}_{+-}^R + \bar{X}^R \bar{\hat{\sigma}} + \alpha \bar{X}^L \bar{\hat{\sigma}} \, ,\\
    0 & = \alpha M_{++}^R - 2 i D_+ X^L - \alpha 2 i D_+ X^R \, ,\\
    0 & = - M_{-+}^L + \bar{\hat{\sigma}} X^L + \alpha \bar{\hat{\sigma}} X^R \, ,\\
    0 & = \alpha M_{--}^L - 2 i D_- X^R - \alpha 2 i D_- X^L \, ,\\
    0 & = - M_{+-}^R + \hat{\sigma} X^R + \alpha \hat{\sigma} X^L \, .
  \end{align}
  Similarly, varying the fields $\eta_-^L$, $\eta_+^R$, $\bar{\eta}_-^L$ and $\bar{\eta}_+^R$, we obtain
  \begin{align}
    0 & = - \bar{\psi}_+^L - \alpha \bar{\psi}_+^R \equiv - \sqrt{\alpha^2 + 1} \bar{\psi}_+^1\, ,\\
    0 & = \bar{\psi}_-^R + \alpha \bar{\psi}_-^L \equiv \sqrt{\alpha^2 + 1} \bar{\psi}_-^1\, ,\\
    0 & = -\psi_+^L - \alpha \psi_+^R \equiv - \sqrt{\alpha^2 + 1} \psi_+^1\, ,\\
    0 & = \psi_-^R + \alpha \psi_-^L \equiv \sqrt{\alpha^2 + 1} \psi_-^1\, .
  \end{align}
  Orthogonal to these fields, we can define
  \begin{align}
    \bar{\psi}_+^2 & \equiv \frac{1}{\sqrt{\alpha^2 + 1}} (\alpha \bar{\psi}_+^L - \bar{\psi}_+^R)\, ,\\
    \bar{\psi}_-^2 & \equiv \frac{1}{\sqrt{\alpha^2 + 1}} (\bar{\psi}_-^L - \alpha \bar{\psi}_-^R)\, ,\\
    \psi_+^2 & \equiv \frac{1}{\sqrt{\alpha^2 + 1}} (\alpha \psi_+^L - \psi_+^R)\, ,\\
    \psi_-^2 & \equiv \frac{1}{\sqrt{\alpha^2 + 1}} (\psi_-^L - \alpha \psi_-^R)\, .
  \end{align}
  We can regard them as the physical fermionic fields. Let us call them $\psi_\pm'$ and $\bar{\psi}_\pm'$.

  Integrating out these auxiliary fields will give us the on-shell Lagrangian consisting of three parts, the kinetic terms for the bosons and fermions, and their interaction,
  \begin{align}\label{eq:Lbos}
    \mathcal{L}_{\textrm{bos}} & = \left(\bar{X}^L\quad \bar{X}^R \right)\cdot
    \left(\begin{array}{cc}
      \square + D + \alpha^2 |\hat{\sigma}|^2 & \frac{1}{\alpha} \square + \alpha D + \alpha |\hat{\sigma}|^2\\
      \frac{1}{\alpha} \square + \alpha D + \alpha |\hat{\sigma}|^2 & \square + D + \alpha^2 |\hat{\sigma}|^2
    \end{array} \right)\cdot
    \left( \begin{array}{c} X^L \\ X^R \end{array} \right) \nonumber\\
    {} & \quad + \bar{F}^L F^L + \bar{F}^R F^R + \alpha \bar{F}^L F^R + \alpha \bar{F}^R F^L\, ,
  \end{align}
  \be\label{eq:Lferm}
     \mathcal{L}_{\textrm{ferm}} = -\frac{\alpha^2 - 1}{\alpha^2 + 1} \bar{\psi}_-' 2 i D_+ \psi_-' - \frac{\alpha^2 - 1}{\alpha^2 + 1} \bar{\psi}_+' 2 i D_- \psi_+' - \chi_-^L 2 i D_+ \bar{\chi}_-^L - \chi_+^R 2 i D_- \bar{\chi}_+^R\, ,
  \ee
  \begin{align}\label{eq:Lint}
    \mathcal{L}_{\textrm{int}} & = - \bar{\psi}_-^L \hat{\sigma} \psi_+^L - \bar{\psi}_+^L \bar{\hat{\sigma}} \psi_-^L + \bar{X}^L i (\lambda \psi^L) - i (\bar{\psi}^L \bar{\lambda}) X^L + \bar{X}^L i \bar{\lambda}_+ \chi_-^L - \bar{\chi}_-^L i \lambda_+ X^L \nonumber\\
    {} & \quad - \bar{\psi}_-^R \hat{\sigma} \psi_+^R - \bar{\psi}_+^R \bar{\hat{\sigma}} \psi_-^R + \bar{X}^R i (\lambda \psi^R) - i (\bar{\psi}^R \bar{\lambda}) X^R - \bar{X}^R i \bar{\lambda}_- \chi_+^R + \bar{\chi}_+^R i \lambda_- X^R \nonumber\\
    {} & \quad - \alpha \bar{\psi}_-^L \hat{\sigma} \psi_+^R - \alpha \bar{\psi}_+^L \bar{\hat{\sigma}} \psi_-^R + \alpha \bar{X}^L i (\lambda \psi^R) - \alpha i (\bar{\psi}^L \bar{\lambda}) X^R + \alpha \bar{\chi}_-^L \bar{\hat{\sigma}} \chi_+^R - \alpha \bar{X}^L i \bar{\lambda}_- \chi_+^R - \alpha \bar{\chi}_-^L i \lambda_+ X^R \nonumber\\
    {} & \quad - \alpha \bar{\psi}_-^R \hat{\sigma} \psi_+^L - \alpha \bar{\psi}_+^R \bar{\hat{\sigma}} \psi_-^L + \alpha \bar{X}^R i (\lambda \psi^L) - \alpha i (\bar{\psi}^R \bar{\lambda}) X^L + \alpha \bar{\chi}_+^R \hat{\sigma} \chi_-^L + \alpha \bar{X}^R i \bar{\lambda}_+ \chi_-^L + \alpha \bar{\chi}_+^R i \lambda_- X^L \, .
  \end{align}

\section{Semichiral St\"uckelberg field}\label{App:St}

Expanding the Lagrangian of the semichiral St\"uckelberg field \eqref{eq:Lst} in components, we obtain
\begin{align}
  {\cal L}_{St} & = -4 (D_+ D_- X_L) (X_L + \bar{X}_L) - 4 (D_- X_L) (D_+ X_L) - \bar{M}_{-+}^L M_{-+}^L + F_L \bar{F}_L \nonumber\\
  {} & \quad + 2 i (D_+ M_{--}^L) (X_L + \bar{X}_L) + M_{--}^L 2 i (D_+ X_L) - \bar{M}_{--}^L 2 i (D_+ X_L) \nonumber\\
  {} & \quad + \bar{\psi}_+^L 2 i (D_- \psi_+^L) - \bar{\psi}_-^L 2 i (D_+ \psi_-^L) + \chi_-^L 2 i (D_+ \bar{\chi}_-^L) - \bar{\eta}_-^L \bar{\psi}_+^L - \eta_-^L \psi_+^L \nonumber\\
  {} & \quad + 2 i D^0 r_L^0 + i \lambda_-^0 \bar{\psi}_+^{L0} - i \bar{\lambda}_-^0 \psi_+^{L0} \nonumber\\
  {} & \quad -4 (D_- D_+ X_R) (X_R + \bar{X}_R) - 4 (D_- X_R) (D_+ X_R) - \bar{M}_{+-}^R M_{+-}^R + F_R \bar{F}_R \nonumber\\
  {} & \quad - 2 i (D_- M_{++}^R) (X_R + \bar{X}_R) - 2 i (D_- X_R) M_{++}^R + 2 i (D_- X_R) \bar{M}_{++}^R \nonumber\\
  {} & \quad - \bar{\psi}_-^R 2 i (D_+ \psi_-^R) + \bar{\psi}_+^R 2 i (D_- \psi_+^R) - \chi_+^R 2 i (D_- \bar{\chi}_+^R) + \bar{\eta}_+^R \bar{\psi}_-^R + \eta_+^R \psi_-^R \nonumber\\
  {} & \quad + 2 i D^0 r_R^0 + i \lambda_-^0 \bar{\psi}_+^{R0} - i \bar{\lambda}_-^0 \psi_+^{R0} \nonumber\\
  {} & \quad - 4 \alpha (D_+ D_- X_L) (X_L + \bar{X}_R) + \alpha (2 i D_- X_L + M_{--}^L) (2 i D_+ X_L + \bar{M}_{++}^R) \nonumber\\
  {} & \quad + 2 i \alpha (D_+ M_{--}^L) (X_L + \bar{X}_R) + \alpha F_L F_R \nonumber\\
  {} & \quad + \alpha \bar{\psi}_+^R 2 i (D_- \psi_+^L) - \alpha \bar{\psi}_-^R 2 i (D_+ \psi_-^L) - \alpha \bar{\eta}_-^L \bar{\psi}_+^R - \alpha \psi_-^L \eta_+^R \nonumber\\
  {} & \quad + i \alpha D^0 (X_L + \bar{X}_R)^0 + i \alpha \lambda_-^0 \bar{\psi}_+^{R0} \nonumber\\
  {} & \quad - 4 \alpha (D_- D_+ X_R) (X_R + \bar{X}_L) + \alpha (2 i D_- X_R - \bar{M}_{--}^L) (2 i D_+ X_R - M_{++}^R) \nonumber\\
  {} & \quad - 2 i \alpha (D_- M_{++}^R) (X_R + \bar{X}_L) + \alpha F_R \bar{F}_L \nonumber\\
  {} & \quad - \alpha \bar{\psi}_-^L 2 i (D_+ \psi^R) + \alpha \bar{\psi}_+^L 2 i (D_- \psi_+^R) - \alpha  \bar{\psi}_-^L \bar{\eta}_+^R - \alpha \eta_-^L \psi_+^R \nonumber\\
  {} & \quad + i \alpha D^0 (X_R + \bar{X}_L)^0 + i \alpha \lambda_-^0 \bar{\psi}_+^{L0}\, .
\end{align}
where $r_{L, R}$ stand for the real part of $\mathbb{X}_2^{L, R}$, and the upper index $0$ denotes the zero mode. Varying the fields $M_{--}^L$, $\bar{M}_{--}^L$, $M_{++}^R$ and $\bar{M}_{++}^R$, we obtain
\begin{align}
  0 & = - 2 i D_+ \bar{X}_L - 2 i \alpha D_+ \bar{X}_R + \alpha \bar{M}_{++}^R\, ,\nonumber\\
  0 & = - 2 i D_+ X_L - 2 i \alpha D_+ X_R + \alpha M_{++}^R\, ,\nonumber\\
  0 & = 2 i D_- \bar{X}_R + 2 i \alpha D_- \bar{X}_L + \alpha \bar{M}_{--}^L\, ,\nonumber\\
  0 & = 2 i D_- X_R + 2 i \alpha D_- X_L + \alpha M_{--}^L\, .
\end{align}
Similarly, varying the fields $\eta_-^L$, $\bar{\eta}_-^L$, $\eta_+^R$ and $\bar{\eta}_+^R$ will give us
\begin{align}
  0 & = - \psi_+^L - \alpha \psi_+^R \equiv - \sqrt{1+\alpha^2} \psi_+^1 \, ,\nonumber\\
  0 & = - \bar{\psi}_+^L - \alpha \bar{\psi}_+^R \equiv - \sqrt{1+\alpha^2} \bar{\psi}_+^1\, ,\nonumber\\
  0 & = - \psi_-^R - \alpha \psi_-^L \equiv - \sqrt{1+\alpha^2} \psi_-^1\, ,\nonumber\\
  0 & = \bar{\psi}_-^R + \alpha \bar{\psi}_-^L \equiv - \sqrt{1+\alpha^2} \bar{\psi}_-^1\, .
\end{align}
We can define
\begin{align}
  \psi_+^2 & \equiv \frac{1}{\sqrt{1+\alpha^2}} \psi_+^L - \alpha \psi_+^R\, ,\nonumber\\
  \bar{\psi}_+^2 & \equiv \frac{1}{\sqrt{1+\alpha^2}} \bar{\psi}_+^L - \alpha \bar{\psi}_+^R\, ,\nonumber\\
  \psi_-^2 & \equiv \frac{1}{\sqrt{1+\alpha^2}} \psi_-^R - \alpha \psi_-^L\, ,\nonumber\\
  \bar{\psi}_-^2 & \equiv \frac{1}{\sqrt{1+\alpha^2}} \bar{\psi}_-^R - \alpha \bar{\psi}_-^L\, .
\end{align}
Integrating out the auxiliary fields, we obtain
\begin{align}
  {\cal L}_{St} & =
  \left( \begin{array}{cc}
    \bar{X}_L & \bar{X}_R
  \end{array} \right)
  \left( \begin{array}{cc}
    \square & \frac{1}{\alpha} \square \\
    \frac{1}{\alpha} \square & \square
  \end{array} \right)
  \left( \begin{array}{c}
    X_L \\
    X_R
  \end{array} \right) \nonumber\\
  {} & \quad + \frac{i}{2} (\frac{1}{\alpha^2} - \alpha^2) \bar{\psi}_+^2 D_- \psi_+^2 - \frac{i}{2} (\frac{1}{\alpha^2} - \alpha^2) \bar{\psi}_-^2 D_+ \psi_-^2 + \bar{\chi}_-^L 2 i D_+ \chi_-^L - \bar{\chi}_+^R 2 i D_- \chi_+^R \nonumber\\
  {} & = \frac{\alpha - 1}{\alpha} \, \bar{X}_1 \square X_1 + \frac{\alpha + 1}{\alpha} \, \bar{X}_2 \square X_2 \nonumber\\
  {} & \quad + \frac{i}{2} (\frac{1}{\alpha^2} - \alpha^2) \bar{\psi}_+^2 D_- \psi_+^2 - \frac{i}{2} (\frac{1}{\alpha^2} - \alpha^2) \bar{\psi}_-^2 D_+ \psi_-^2 + \bar{\chi}_-^L 2 i D_+ \chi_-^L - \bar{\chi}_+^R 2 i D_- \chi_+^R \nonumber\\
  {} & = \frac{\alpha - 1}{\alpha} \, (\bar{r}_1 \square r_1 + \bar{\gamma}_1 \square \gamma_1) + \frac{\alpha + 1}{\alpha} \, (\bar{r}_2 \square r_2 + \bar{\gamma}_2 \square \gamma_2) \nonumber\\
  {} & \quad + \frac{i}{2} (\frac{1}{\alpha^2} - \alpha^2) \bar{\psi}_+^2 D_- \psi_+^2 - \frac{i}{2} (\frac{1}{\alpha^2} - \alpha^2) \bar{\psi}_-^2 D_+ \psi_-^2 + \bar{\chi}_-^L 2 i D_+ \chi_-^L - \bar{\chi}_+^R 2 i D_- \chi_+^R\, ,
\end{align}
where
\be
  X_1 \equiv \frac{- X_L + X_R}{\sqrt{2}}\, ,\quad X_2 \equiv \frac{X_L + X_R}{\sqrt{2}}\, ,
\ee
while $r_{1, 2}$ and $\gamma_{1, 2}$ denote the real parts and the imaginary parts of $X_{1, 2}$ respectively. Among these real components only one of them, $r_2$, transforms under the gauge transformations.

\section{Jeffrey-Kirwan Residue}\label{App:JK}

In the computation of section~\ref{sec:EG}, we need the Jeffrey-Kirwan residue, which was discussed in Refs.~\cite{JK-1, JK-2}. Here we give a brief discussion following ~\cite{Benini-genus-1, Benini-genus-2, Lee} and the references therein.

Suppose $n$ hyperplanes intersect at $u_* = 0 \in \mathbb{C}^r$, which are given by
\be
  H_i = \{u \in \mathbb{C}^r | Q_i (u) = 0 \}\, ,
\ee
where  $i = 1, \cdots, n$ and $Q_i \in (\mathbb{R}^r)^*$. In the GLSM, $Q_i$ correspond to the charges, and they define the hyperplanes as well as their orientations. Then for a vector $\eta \in (\mathbb{R}^r)^*$, the Jeffrey-Kirwan residue is defined as
\be
  \textrm{JK-Res}_{u=0} (Q_*, \eta)\, \frac{d Q_{j_1} (u)}{Q_{j_1} (u)} \wedge \cdots \wedge \frac{d Q_{j_r} (u)}{Q_{j_r} (u)} = \Bigg\{
  \begin{array}{ll}
    \textrm{sign}\, \textrm{det} (Q_{j_1} \cdots Q_{j_r})\, , & \textrm{if } \eta \in \textrm{Cone} (Q_{j_1} \cdots Q_{j_r}) \\
    0\, , & \textrm{otherwise}\, ,
  \end{array}
\ee
where $Q_* = Q (u_*)$, and $\textrm{Cone} (Q_{j_1} \cdots Q_{j_r})$ denotes the cone spanned by the vectors $Q_{j_1},\cdots , Q_{j_r}$. For instance, for the case $r = 1$,
\be
  \textrm{JK-Res}_{u = 0} (\{ q\}, \eta) \, \frac{du}{u} = \bigg\{
  \begin{array}{ll}
    \textrm{sign} (q)\, , & \textrm{if } \eta q > 0\, ,\\
    0\, , & \textrm{if } \eta q < 0\, .
  \end{array}
\ee

To obtain the elliptic genus, we still have to evaluate the contour integral over $u$. Since in the paper we often encounter the function $\vartheta_1 (\tau, u)$, its residue is very useful in practice:
\be
  \frac{1}{2 \pi i} \oint_{u = a + b \tau} du\, \frac{1}{\vartheta_1 (\tau, u)} = \frac{(-1)^{a+b} \, e^{i\pi b^2 \tau}}{2 \pi \, \eta(q)^3}\, ,
\ee
where $q = e^{2 \pi i \tau}$. This relation can be derived by combining the properties
\be
  \vartheta'_1 (\tau, 0) = 2 \pi \, \eta(q)^3\, ,
\ee
and
\be
  \vartheta_1 (\tau, u + a + b \tau) = (-1)^{a+b}\, e^{-2 \pi i b u - i \pi b^2 \tau} \vartheta_1(\tau, u)
\ee
for $a, b\in \mathbb{Z}$ and the fact that $\vartheta_1 (\tau, u)$ has only simple zeros at $u = \mathbb{Z} + \tau \mathbb{Z}$ but no poles.

\bibliographystyle{JHEP}
\bibliography{Semichiral}

\end{document}